\documentclass[letterpaper,twocolumn,10pt]{article}
\usepackage{usenix2019_v3}
\usepackage[utf8]{inputenc}

\DeclareUnicodeCharacter{FF0C}{,}
\usepackage[linesnumbered,ruled,vlined]{algorithm2e}
\usepackage[noend]{algpseudocode}
\usepackage{array}
\newcolumntype{x}[1]{>{\centering\arraybackslash\hspace{0pt}}p{#1}}
\usepackage{amsmath}
\usepackage{amssymb}
\usepackage[inline]{enumitem}
\usepackage{epsfig}
\usepackage{fancyvrb}
\usepackage{graphicx}
\usepackage{booktabs, tabularx}
\usepackage[flushleft]{threeparttable}
\usepackage{latexsym, color}
\usepackage{listings}
\usepackage{tikz}
\usepackage{url}
\usepackage[font=normalsize,labelfont=bf]{caption}
\usepackage{hyperref}
\usepackage{cleveref}
\usepackage{tabularx}
\usepackage{subcaption}
\usepackage{xcolor}
\usepackage{multirow}
\usepackage{multicol}
\usepackage{float}
\usepackage{array}
\usepackage{setspace}
\usepackage{stfloats}
\usepackage{arydshln}
\usepackage{adjustbox}  % 导言区添加，用于灵活调整图片尺寸

\crefformat{figure}{Figure~#2#1#3}
\definecolor{commentgreen}{RGB}{2,112,10}
\definecolor{eminence}{RGB}{108,48,130}
\definecolor{weborange}{RGB}{255,165,0}
\definecolor{frenchplum}{RGB}{129,20,83}

\long\def\comment#1{}

\usepackage{cuted}
\usepackage{float}
\usepackage{xspace}

\usepackage{amssymb}

\newcommand{\groot}{GR\textsc{oot}}
\newcommand{\system}{Janus}

\newcommand{\eg}{{\em e.g.}}

\newcommand{\para}[1]{\noindent {\bf #1}}
\usepackage{fancyhdr}
\newtheorem{defn}{Definition}

\begin{document}

\title{\system{}: Leveraging Incremental Computation for Efficient DNS Verification}
%
% Aether: An Efficient Incremental DNS Configuration Verification Framework

%only change numbers when submitting

\author{Yao Wang$^{\dagger}$,
Kexin Yu$^{\dagger}$,
Wenyun Xu$^{\dagger}$,
Ziyi Wang$^{\dagger}$,
Kaiqiang Hu$^{\star}$,
Lizhao You$^{\dagger}$,
Qiang Su$^{\dagger}$,\\
Dong Guo $^{\diamondsuit}$,
Haizhou Du$^{\star}$,
Wanjian Feng$^{\triangle}$, 
Qingyu Song$^{\dagger}$,
Linghe Kong$^{\ddagger}$,
Qiao Xiang$^{\dagger}$,
Jiwu Shu$^\blacktriangle$$^{\dagger}$
\vspace{0.3cm} \\
$^\dagger$Xiamen University,
$^{\star}$Shanghai University of Electric Power,
$^{\triangle}$Yealink Technologies Co. Ltd., China,\\
$^{\diamondsuit}$Fudan University,
$^{\ddagger}$Shanghai Jiao Tong University, ,$^\blacktriangle$Minjiang University
}

%\renewcommand{\shortauthors}{Anonymous Author, et al.}

%\ifcase \conftype 
%\input{abstract}
%\maketitle

%\or 
\maketitle
\begin{abstract}
Existing DNS configuration verification tools face significant issues (\eg, inefficient and lacking support for incremental verification).
%DNS configuration verification tools still face significant issues in leading to much redundant response behavior and cannot support incremental update.
%
% And current research still lacks effective automated solutions for diagnosing and repairing DNS misconfiguration. 
Inspired by the advancements in recent work of distributed data plane verification and 
% the control plane diagnosis and repair
% and 
the resemblance between the data plane and DNS configuration, we tackle the challenge of DNS misconfiguration by introducing \system{}, a DNS verification tool. Our key insight is that the process of a nameserver handling queries can be transformed into a matching process on a match-action table. With this insight, \system{} consists of (1) an efficient data structure for partition query space based on the behaviors, (2) a symbolic execution algorithm that specifies how a single nameserver can efficiently cover all possible queries and ensure the accuracy of verification, (3) a mechanism to support incremental verification with less computational effort. Extensive experiments on real-world datasets (with over 6 million resource records) show that \system{} achieves significant speedups, with peak improvements of up to 255.7× and a maximum 6046× reduction in the number of LECs.

% \system{} also achieves a speedup in incremental update scenario for 99.99\% of zones, with a maximum speedup of 3,370$\times$.

%Centralized DNS misconfiguration verification faces significant scalability issues in large-scale DNS system (e.g., the performance bottleneck and not support incremental verification). Inspired by the recent work of distributed data plane verification and the resemblance between the data plane and DNS configuration, we tackle the scalability challenge by introducing ~\system{}, a distributed DNS verification framework. Our key insight is that by analyzing the behavior of a single identity (e.g., server, zonefile, process) and parallelizing the computation of DNS configuration verification, we can substantially scale up the verification process. With this insight, Octopus consists of (1) an invariant specification language, (2) a novel data structure Action Tire to store and compute local equivalence class, (3) a distributed DNS configuration verification protocol that specifies how single identities efficiently
%communicate task results to jointly verify the invariants, and (4) a mechanism to support incremental verification with less computational effort.
%Extensive experiments with real-world datasets show that \system{} can achieve an up to xxxx speed up on a dataset with over 410,000 resource records while having small overhead.
 % Evaluation shows that the design is scalable in real networks while having low overhead.

\end{abstract}

%\fi

 \section{Introduction}
\label{intro}
Over the past three decades since the initial publication of two RFCs~\cite{rfc1034, rfc1035} defining the basic Domain Name System (DNS), the DNS system has evolved into a crucial component of the Internet. As one of the largest distributed hierarchical database systems, its immense scale and complexity pose significant challenges for configuration~\cite{dnsdev, pang2004, schomp2013, shaikh2001,rfc1035,rfc1912}. Any failure in the DNS could lead to incalculable losses~\cite{Cloudfare, miscase1, miscase2, miscase3, miscase4, miscase5}.

\para{Static methods support limited properties.}
To ensure uninterrupted DNS operations, operators employ various techniques, primarily centered on probing (\eg, linting and testing)~\cite{monitoring_zdrnja, checkhost2020, MXToolbox, dig, dnsrecon, microsoft2020dnslint} and the static analysis of individual files~\cite{xu_2015_systems, namedcheck, bind9}. However, these tools have limited support for verifiable properties and lack precision. For example, the linting-based approach focuses on detecting basic syntax and formatting issues, and may not be able to detect complex logical errors or context-related problems. 
%The testing-based approach has the situation that test results may be affected by changes in the environment, for example, the different DNS server configurations. 

%The research related to DNS verification has been persistently underway for a significant duration of time. Early efforts %primarily 
%mainly focused on probing~\cite{monitoring_zdrnja, checkhost2020} and static analysis of single files~\cite{xu_2015_systems, namedcheck,bind9}. In recent years, with the application of formal methods in network verification~\cite{formal1, formal2, formal3, formal4, formal5, symbolic}, tools like {\groot}~\cite{kakarla2020groot} %have begun 
%began to focus on formally defining the DNS system %for 
%by validating a set of zonefiles. It scrutinizes DNS zonefiles against specified properties, verifying whether these properties hold for all potential DNS queries or providing a counterexample if they do not.
\para{Existing DNS configuration verification tools suffer from unsound definitions of Equivalence Class (EC), inefficient performance, and the lack of support for incremental update.}
Some studies~\cite{kakarla2020groot,liu2023formal,Octopus} attempt to tackle the limitations of static analysis methods. \groot{}~\cite{kakarla2020groot} uses a centralized architecture to collect zonefiles are then passed to a centralized verifier for unified verification.
% It categorizes all possible queries into equivalence classes (EC) and performs symbolic queries on each class to minimize computational complexity and explore all possible DNS queries or provide a counterexample.
Octopus~\cite{Octopus} proposes a distributed DNS verification framework that allows each authoritative DNS server in the DNS system to focus on its own managed zonefiles and conduct behavioral analysis on each zonefile of the server. However, they still suffer from some issues.

First of all, we execute the source code for \groot{} and observed certain scalability performance limitations. For 1,000,000 zones, it takes \groot{} about 8455.13s to finish the verification. These findings indicate that \groot{} demonstrates inefficiency even with moderate dataset sizes, which could impact its practicality in scenarios requiring faster verification speeds.
% Such variations indicate that \groot{}'s ECs are unsound with respect to the semantic model, as queries that should be treated similarly can result in different outcomes in resolution.
% there exists the situation that two different queries that are parsed in the same way in \groot. They give an example involves two zones, example.com and \url{example.net}, each with different DNS records. When querying \url{a.example.com}, the resolver encounters a rewrite loop and signals an error, while querying \url{b.example.com} leads to a different resolution process that fails to detect the loop until another query is made. 
% This is because the queries are not in the zonefile's record, they will be matched against the wildcard without any further distinction between the namespaces contained in $\alpha$, but it is possible that they point to different name spaces during resolution. 
%Second, \groot{} needs to collect the zonefiles from various DNS servers to a trusted centralized server. This is hard for a decentralized system such as DNS. The formal method~\cite{liu2023formal} has developed a formal end-to-end semantic name resolution for the formal analysis of qualitative and quantitative properties, and has proposed an automated tool to detect DoS attacks. It employs random zone generation and heuristic coverage, which can effectively identify issues in DNS features, however, this approach has a limited search space.
Secondly, the EC definition of \groot{} is unsound. As mentioned in the prior study~\cite{liu2023formal} the same EC can exhibit different resolution behaviors in \groot{}. This inconsistency arises when the same type of query leads to one being resolved with an immediate detection of a rewrite loop, while another requires additional queries to identify the same loop. Lastly,  \groot{} is intended to provide global verification of the entire DNS configuration. When the operator performs an incremental update of the DNS configuration, the logical structure of the algorithm may not support independent verification of only a portion of the configuration and can only update the entire domain name space, which does not support incremental update. Octopus ~\cite{Octopus} plans to support incremental verification, but it has not yet been implemented.
In this paper, we systematically tackle the important problem of DNS verification tools to be efficient, complete, and support incremental verification in the DNS ecosystem. 
%%Not only can an efficient DNS verification tool quickly find errors, but it can also shed light on the way forward for new research topics related to DNS, such as the diagnosis and repair of DNS configurations.
% In this paper, we systematically tackle the important problem of DNS verification, making it possible to verify configuration files in a scalable and automated way for large-scale DNS ecosystems. The scalable DNS life cycle verification tool can quickly find errors in large-scale DNS ecosystems and suggest modifications for them, which greatly improves the efficiency of operation and maintenance personnel.

\para{Proposal: Design an algorithm to compress the number of ECs, thereby reducing processing overhead.}
Inspired by the studies of data plane verification on ECs~\cite{ec1, ec2, ec3, ec4, ec5, ec6, anderson2014netkat}, through a more in-depth analysis of the behavior of the nameserver, we compress ECs to reduce their number. A smaller number of ECs means that the volume of data to be processed can be reduced, thereby accelerating the speed of configuration verification. Additionally, we implement an efficient data structure to store and compute the EC and develop an incremental update algorithm based on the new EC definition.

\para{Key insight: Nameserver can be modeled as a match-action table.}
The fundamental challenge in realizing DNS verification including:
(1) how to construct and store each rule in the match-action table, (2) how to verify the configuration files through the query matching process,  
% While the former study Octopus~\cite{Octopus} presents the promise of distributed DNS verification, it fails to address several important questions: %(1) how to effectively specify and verify invariants, 
% (1) how to store and compute LEC, %(3) how to minimize the information exchange between identities to reduce overhead,
and (3) how to adapt to the frequent changes in configurations within the DNS system. To this end, we design \system{}, a generic DNS configuration verification framework, with a key insight: by analyzing the query processing behaviors of nameservers to construct match-action tables and stitching the results in a symbolic way, we can substantially scale up the verification of DNS configuration. \system{} has 3 key designs (D1-D3):

%\para{D1: An invariant specification language (\S \ref{sec: specification}).} This language abstracts invariants into situations and clues, where situations are simple modelling of system execution and clues are constraints on execution. It allows operators to flexibly specify commom invariants studied by existing DNS configuration verification tools (\eg{}, Missing Glue Records, Rewriting Blackholing and Answer Inconsistency). In addition, we provide a template to help you represent the implemented model more quickly.
% \begin{figure}[ht!]
%     \centering
%     \includegraphics[width=1\linewidth]{figures/architecture.pdf}
%     \caption{The architecture and workflow of \system{}.}
%     \label{fig:workflow_withoutexample}
% \end{figure}
\para{D1: A novel data structure to encode query space in each match-action table rule (\S\ref{sec:Query Space Partitioning}).} We use an equivalence class (EC) to aggregate DNS queries with the same behavior. In pursuit of the smallest possible number of ECs, we introduce a novel data structure and corresponding encode algorithm to calculate ECs. By aggregating all queries with the same behaviour, the ECs are disjointed from each other to obtain the minimum.

\para{D2: A symbolic query algorithm to cover all possible query spaces (\S\ref{sec:Symbolic Query Execution}).} To achieve comprehensive coverage of all query possibilities, we utilize symbolic execution, a technique that enables us to explore various query paths and behaviors in \system{} without executing the actual query. We design a symbolic execution algorithm that takes a set of nameservers, a query, and a fuel parameter to simulate DNS behavior. By analyzing different resolution behaviors, the algorithm marks responses accordingly, allowing for the classification of various outcomes.

\para{D3: A dynamic update with incremental verification algorithm (\S \ref{sec:Incremental Verification}).} DNS being an extremely large system, its configuration is updated extremely frequently and commonly. We regenerate execution traces affected by configuration updates, focusing on the affected query processing paths while maintaining the integrity of the verification. 

% \para{D4: An algorithm that utilizes formal methods for diagnosis. (\S \ref{sec: diagnosis}).} 
%  Specifically, the algorithm converts the trace information obtained during the verification process into Satisfiability Modulo Theories (SMT) formulas. Through this conversion, complex configuration issues are transformed into mathematical logic problems, thereby leveraging the reasoning capabilities of SMT solvers to detect errors. The algorithm also employs Minimal Unsatisfiable Core (MUC) analysis techniques to automatically identify and locate the specific records causing configuration errors. MUC analysis helps to precisely determine which configuration items are the root causes of the errors by gradually narrowing down the set of unsatisfiable constraints.

% \para{D5: A repair algorithm with incremental scope expansion and intelligent modification. (\S \ref{sec:repair}).} 
% As a general framework based on a template-based synthesis method, the algorithm gradually expands the repair scope and intelligently defines the value space required for modifications. The algorithm ensures that the repair results comply with the specifications while preserving the correctness of other execution traces. Once generated, the repair suggestions are submitted to administrators for confirmation. After approval, the suggestions are sent back to the verifier for incremental verification, thereby effectively improving the accuracy and efficiency of the repair process.

\para{Evaluation.}
We implement the prototype of \system{} and evaluate it with a real-world open-source dataset~\cite{dnscensus2013} and a university dataset. We compare \system{} with the first DNS verification tool: \groot{}. For the open-source data set, we evaluate the end-to-end verification time that concludes the EC construction time, symbolic execution time, and incremental time. %It explains, the reason for the smaller verification time of \system{}. 
 Additionally, the maximum number of {\groot} end-to-end verifications can be as high as 255.7$\times$ that of \system{}. In the incremental update scenario, we achieve a maximum speedup of 3,370$\times$.

 \begin{figure*}[htbp]
    \centering
    % Fig1 单独一行
        \includegraphics[width=\textwidth]{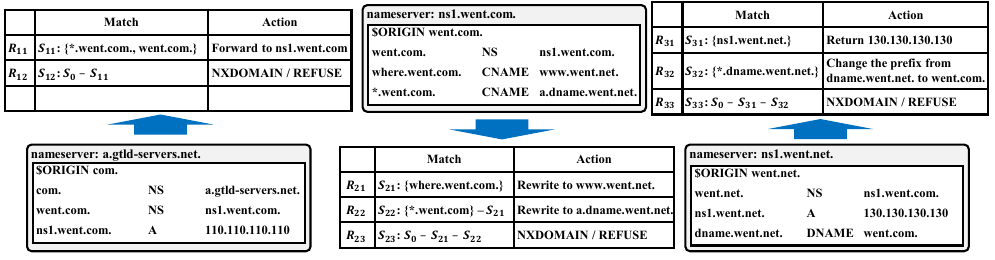}
        %\label{fig: wf.ct}
   
    \caption{Example zonefiles for three nameservers and the result of query space partitioning for each nameserver. For simplicity, we merge the NXDOMAIN and REFUSE behaviors. $S_0$ denotes the entire query space.}
    \label{fig: workflow}
\end{figure*}

\section{Overview}
This section introduces some key concepts of \system{}, and illustrates its workflow using an example.
%\vspace{-1mm}
\subsection{Basic Concept}
\para{Simplification of the DNS System.} Generally, the DNS system comprises three main roles~\cite{rfc3467}: clients, responsible for initiating and receiving query requests; resolvers, responsible for sending query requests to nameservers and receiving responses to serve the clients; and nameservers, responsible for storing the mappings between domain names and IP addresses to respond to resolver queries. In practice, these components can be further subdivided, such as into recursive resolvers and iterative resolvers.

To simplify verification challenges, we focus on the behavior of the nameservers that store configuration files, akin to the approach adopted in \groot{} \cite{kakarla2020groot}. A nameserver typically consists of one or more configuration files, also referred to as zonefiles. Each zonefile contains a series of resource records (RRs) that describe the mappings between the domain names managed by the server and their corresponding IP addresses.

Each resource record (RR) comprises five components: domain name, TTL, class, type, and data. The domain name specifies the domain associated with the resource record (\eg, www.google.com). The TTL (Time to Live) indicates the lifespan of the resource record. The class denotes the type of network the record belongs to (typically IN for Internet). The type specifies the data type of the resource record (\eg, NS (designate the authoritative DNS server for a specific domain name), A (IPV4 address), AAAA  (IPV6 address), CNAME (map one domain name to another domain name), etc.), and the data represents the content of the record (\eg, an IP address or another domain name).

Since the class attribute is rarely used outside of IN, it will not be considered in subsequent discussions. Although TTL is a critical attribute of RRs, particularly for resolver caching, our focus on nameserver configuration files allows us to exclude TTL from further consideration. Errors related to TTL can be verified using existing tools such as named-checkzone \cite{namedcheck}. Consequently, we simplify the RR to three components: $\langle \textit{rname}, \textit{rtype}, \textit{rdata}\rangle$, where $\textit{rname}$ denotes the domain name, $\textit{rtype}$ represents the record type, and $\textit{rdata}$ denotes the record's data. And the 

Furthermore, other DNS mechanisms, such as DNS caching and DNS Security Extensions (DNSSEC) \cite{Roll}, are excluded from consideration in this study to simplify the problem. In future work, we will progressively extend our model to accommodate additional DNS mechanisms.

\para{DNS Model.} To facilitate explanation, a domain nameserver can be modeled as a match-action table, where each row represents a rule. Each rule consists of two parts: matching conditions and actions. When a query request $\langle \textit{qname}, \textit{qtype}\rangle$ arrives, it is matched against each rule in the table. Upon a successful match, the corresponding action is executed. The possible actions can be broadly categorized into two types: terminate and forward. A terminate action indicates that, upon a successful match, the query should terminate, returning the corresponding result (\eg, the resolved data) or signaling a query failure. A forward action indicates that, upon a successful match, the query should be forwarded to the next domain nameservers for further resolution. The next servers could either be the current server itself or a different domain nameserver. Similar to the approach in \groot{}, we assume that all configuration files required for verification are available and explicitly define which configuration files belong to each domain nameserver.

%我们将LEC构建这一过程抽象为BDD的构建。首先我们将所有record分为三类：NS记录，DNAME记录以及other记录（我们将除了NS记录和DNAME记录外的其他记录的优先级视为相同）。在抽象为BDD的过程中，我们首先计算label集合：label_set。我们要将zonefile中的所有label提取出来并进行标号（对于DNAME记录和CNAME记录的rdata我们认为他们也是 一个域名，我们还回额外增加两个label来代表空集和other），以此来计算label_bit即BDD中变量的数量,其中 label_bit=xxxx，当我们获得label_bit后，就可以根据变量来表示等价类所代表的空间。\\
\para{Query Log and Traces.} %Inspired by  Tulkun~\cite{tulkun}, 
We introduce \emph{query trace} to record the processing workflow of a query, thereby defining the behavior of a query within the DNS system. When a query \( q \) passes through a DNS system, each successful match against a rule is defined as a query log. A log captures information about the initial value of the query, its final value (if it has been rewritten), and the location where the match occurred.

A query trace for \( q \) is then defined as an ordered sequence of logs, where the action in the final log must always be of the terminate type. However, for actions of the forward type, the number of subsequent domain nameservers cannot be deterministically identified. For example, it is well-known that root nameservers consist of 13 groups. A query \( q \) initiated by a user will be forwarded simultaneously to all 13 groups of root nameservers. As a result, a single query \( q \) may correspond to multiple query traces.

The concept of traces serves as the foundation for verification. All properties that require verification will be validated across all possible traces.

\subsection{Workflow}

We demonstrate \system{}’s workflow with the example in Figure~\ref{fig: workflow} with the input query: $q_{origin}:\langle \text{*.com.}, \text{A} \rangle$. The properties we verify here are that no query in the system, after being rewritten, fails to obtain a result (rewrite blackholing) or causes a loop (rewrite loop). We will also demonstrate how to perform incremental verification after fixing rewrite blackholing. In order to facilitate comprehension and streamline the demonstration process, this section assumes the presence of a single query of type A, thus omitting the type field from the query. Consequently, the input request will be denoted by $q_{origin}$:\{ \text{*.com.}\}.

% \vspace{-1mm}
\subsubsection{Query Space Partition}
Given a domain nameserver, \system{} partitions the entire query space into several subspaces. After partitioning, the corresponding action for each subspace is determined. Each subspace can then be treated as a matching rule. Essentially, this process computes the match-action table. The core challenge lies in ensuring the completeness and mutual exclusivity of the partitioned subspaces. Specifically:  

1. $\bf{Completeness}$ ensures that every possible query request can match exactly one subspace, guaranteeing the correctness of query processing.

2. $\bf{Mutual\ Exclusivity}$ ensures that no two subspaces overlap, thereby ensuring that a given query corresponds to a unique action.  

To address this challenge, each subspace is treated as a set, and the problem is formulated as a set coverage problem, which can be solved using a greedy algorithm. The detailed algorithm will be elaborated in Section \ref{sec:Query Space Partitioning}.  

Figure \ref{fig: workflow} illustrates an example of computing a match-action table. In this example, match-action tables are computed for three domain nameservers. Since the computations for different domain nameservers are independent, they can be executed in parallel.  

Let's take nameserver a.gtld-servers.net as an example and \( S_0 \) denotes the entire query space. For the domain nameserver a.gtld-server.net, the query space is partitioned into two subspaces:  

\( S_{11} \): \{\text{*.went.com.}, \text{went.com.}\}, indicating that queries in this subspace are forwarded to the nameserver ns1.went.com.  

\( S_{12} \): $S_0$ - $S_{11}$, indicating that queries in this subspace could not get the correct response. In fact, we can further partition the space into \{\text{*.com.}, \text{com.}\} - $S_{11}$ and $S_0$ - \{\text{*.com.}, \text{com.}\} according to whether the action is "domain does not exist" (NXDOMAIN) or "refused service" (REFUSE).

The subspaces partitioned for the remaining two domain nameservers are listed in Figure \ref{fig: workflow} and are not elaborated further here.

\subsubsection{Trace Generation and Property Check} 
Once the match-action table is constructed, the process of generating query traces can begin. This process can be conceptualized as a depth-first search (DFS). Given a query \( q_{in} \) and a domain nameserver \( ns \), We first match the query \( q \) with the match-action table of $ns$, which allows us to further refine the behavior of \( q \). In other words, the match-action table partitions \( q \) into a set of subspaces \( \{q_{in1}, q_{in2}, \dots\} \), and retrieves the corresponding action for each subspace \( q_{in} \). If the action is of a termination type, we directly obtain a log. Otherwise, we execute the behavior specified by the action, obtain the query space \( q_{out} \) to be forwarded (which may be the same as \( q_{in} \) or different), record the information to generate a log, and forward \( q_{out} \) as a new request to the next nameserver.
This process continues until the final action is of the terminate type, at which point a complete trace is generated. It is worth noting that due to the forwarding operation, \( q_{out} \) may be forwarded to multiple domain nameservers. As a result, the logs of different traces may overlap. However, this does not affect the definition or validity of individual traces.

Consider the scenario in Figure \ref{fig: wf.se}, where a query \( q_{origin} \) is input to the domain nameserver a.gtld-server.net. We only consider the query matches rule \( R_{11} \), generating a log \( L_1 \), where \( q_{in1} = q_{out1} = S_{11} \). The query \( q_{out1} \) is then forwarded to ns1.went.com.  

At ns1.went.com, we focus on the partial subspaces where the query matches successfully. The first match is with rule \( R_{21} \), which modifies the input query \( q_{in2} \) = \{where.went.com.\} to a new output query \( q_{out2}\) =  \{\text{www.went.net.}\}  and forwards it to ns1.went.net. A log \( L_2 \) is generated for this match. Subsequently, at ns1.went.net, the query \( q_{out2} \) matches rule \( R_{33} \), generating a log \( L_4 \). Here, \( q_{in3} = q_{out3}\) = \{\text{www.went.net.}\} . Since the action corresponding to \( R_{33} \) is of the terminate type, the first trace is formed as: \(
T_1 = [L_1, L_2, L_4]
\).

Next, we consider another match for \( q_{out1} \) at ns1.went.com, this time with rule \( R_{22} \). The action for \( R_{22} \) modifies the query \( q_{in3} = S_{22} \) to \( q_{out3} = \{\text{a.dname.went.net.}\} \) and forwards it to ns1.went.net, generating a log \( L_3 \). At ns1.went.net, the query \( q_{out3} \) matches rule \( R_{33} \), which modifies \( q_{in5} = \{\text{a.dname.went.} \\ \text{net.}\} \) to \( q_{out5} = \{\text{a.went.com.}\} \), generating a log \( L_5 \). The query is then forwarded back to ns1.went.com, where it matches \( R_{23} \) again, generating a log \( L_6 \). Although further forwarding might occur, we stop here because of loop detection and define the second trace as: \(T_2 = [L_1, L_4, L_5, L_6]\).

After obtaining these two traces, we can proceed to verify properties. For a given trace, \system{} examines whether its behavior aligns with the semantics of specified properties.  

For example, the rewrite blackholing property is defined as follows: if a rewrite operation occurs during the query process and the query ultimately fails to retrieve an answer result. Analyzing trace \( T_1 \), we find that a rewrite operation occurs in \( L_2 \), and the final action is NXDOMAIN. Therefore, \( T_1 \) satisfies the rewrite blackholing property.  

Similarly, the rewriting loop property is defined as follows: if a rewrite operation occurs during the query process and the same domain nameserver receives the same query twice within the query path. Analyzing trace \( T_2 \), we observe that rewrite operations occur in both \( L_3 \) and \( L_5 \). Furthermore, the query is forwarded to ns1.went.com in both cases, and the intersection of \( q_{in3} \) and \( q_{in6} \) is \( \{\text{a.went.com}\} \), indicating that ns1.went.com received the same query twice. Therefore, \( T_2 \) satisfies the rewriting loop property. Note that as mentioned in previous work~\cite{liu2023formal}, \groot{} does not have a mechanism to make such a determination, so it is unable to detect the occurrence of loops in a timely manner.

\subsubsection{Incremental Verification}

When configuration files are updated, incremental verification allows us to quickly validate whether the new configuration satisfies the specified properties. This process primarily involves two steps: repartitioning the query space and incrementally generating and verifying traces.  

A straightforward and intuitive approach is to recompute the entire match-action table for domain nameservers with updated configurations. Then, all traces are regenerated from scratch, and properties are reverified. To reduce computation, previously cached traces can be leveraged.  

\begin{figure*}[t]
    \centering
    % Fig1 单独一行
    
    % Fig2和Fig3并排，6:4比例
    \begin{subfigure}{0.60\textwidth}
        \includegraphics[width=\linewidth]{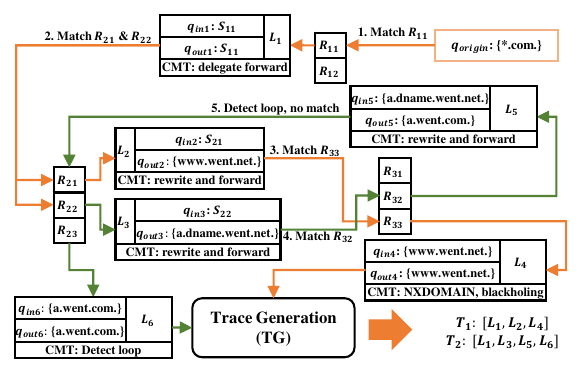}
        \caption{The process of symbolic execution and generating two traces, $T_1$\\ and $T_2$. CMT is a comment on some key information in a log.}
        \label{fig: wf.se}
    \end{subfigure}
    \begin{subfigure}{0.38\textwidth}
        \includegraphics[width=0.9\linewidth]{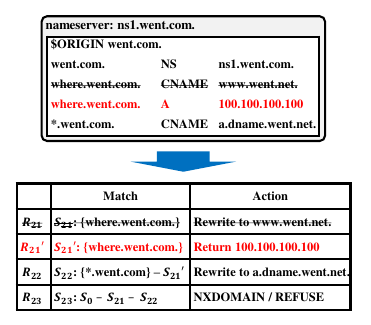}
        % \caption{After changing a CNAME record (strikethrough) in ns1.went.com to an A record (red), the result of the re-partitioned query space.}
        \caption{The result of the re-partitioned query space after changing a CNAME record (strikethrough) in ns1.went.com to an A record (red).}
        \label{fig: wf.ict}
    \end{subfigure}
    
    % Fig4 单独一行
    \begin{subfigure}{0.9\textwidth}
        \includegraphics[width=\textwidth]{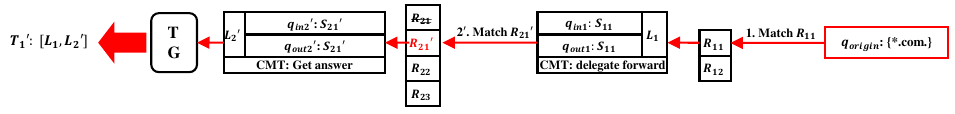}
        \caption{The process of re-executing symbolic execution to update $T_1$ to $T_1'$.}
        \label{fig: wf.ise}
    \end{subfigure}
    
    % \vspace{-5pt}
    \caption{An illustration example in Figure \ref{fig: workflow} to demonstrate the workflow of \system{}.}
    \label{fig: workflow_2}
\end{figure*}

Take the scenario in Figure \ref{fig: wf.ict} as an example. Suppose the record $\langle \text{where.went.com.}, \text{CNAME}, \text{www.went.net.}\rangle$ is repla-\linebreak ced with a new A type record $\langle\text{where.went.com.}, \text{A}, \text{100.100.} \\ \text{100.100}\rangle$. In this case, we only need to recompute the match-action table for nameserver ns1.went.com. Rule \( R_{21} \) is replaced by \( R_{21}' \), whose action directly returns the IPv4 100.100.100.100 and terminate.

Next, we regenerate all traces from scratch. For trace \( T_1 \), we observe that the rewrite operation in \( L_2 \) will no longer occur. Instead, $q_{out1}$ obtains the query result after matching $R_{21}'$, generating a new log $L_2'$. Thus, the trace $T_1$ is replaced by $T_1': [L_1, L_2']$, which means the trace no longer satisfies the rewrite blackholing property.

 %\newpage
 \newcommand{\GetSpace}[3]{\textsc{GetSpace}(#1, #2, #3)}

\section{Query Space Partitioning}\label{sec:Query Space Partitioning}
% 本节我们将介绍每个域名服务器如何构建自己的match-action table，以及如何对其的一些优化操作。
In this section, we will clarify how each nameserver constructs its own match-action table, along with some optimization techniques applied to it.

\subsection{Information Storage}
For each nameserver object, the key data it should store is the LEC (Local Equivalence Class). Given a nameserver $X$, a LEC represents a set of queries whose actions are identical at $X$. $X$ stores its LECs in a $(\textit{query\_space},\textit{aciton})$ mapping called the LEC table (match-action table). We choose to encode query sets as $predicates$ using binary decision diagram (BDD)~\cite{BDD}. This is because the nameserver object performs query set operations (\eg{}, $\cup$ and $\cap$), which can be realized efficiently using logical operations on BDD. 

% 然而我们无法简单的使用前人在数据平面验证中的工作。在数据平面验证时，通常数据包头的长度是固定的（例如如果仅考虑IPv4匹配，那么仅仅需要32比特）。而在DNS中，域名由多个标签（Label）组成，标签之间用点（.)分割。每个标签的长度最大为63个字符，整体域名的最大长度为255个字符。如果直接对其进行编码，我们需要使用255*8=2040个比特。这将使得BDD中的逻辑操作变得极其缓慢。为此我们设计了一个压缩算法1，将域名中的每个label压缩为一个整数，以降低所需的比特数量。
We cannot simply adopt previous work in data plane verification (DPV). In DPV, the length of packet headers is typically fixed (\eg, IPv4 matching only needs 32 bits or variables in BDD). In DNS, domain names consist of multiple labels, separated by dots (.). Each label can be up to 63 characters long, and the overall domain name can be up to 255 characters in length. If we were to directly encode such domain names, we would require 255 * 8 = 2040 variables. This would render logical operations in BDDs exceedingly slow. 
To address this issue, we devise a compression Algorithm \ref{alglabel} that reduces each label in the domain name to an integer, decreasing the number of bits required.

\begin{algorithm}[t]
    \footnotesize
    \caption{Encode labels as integer}\label{alglabel}
    
    \SetKwFunction{UpdateEncodeMap}{UpdateEncodeMap}
    \SetKwFunction{EncodeLabels}{EncodeLabels}
    \SetKwProg{Fn}{Function}{:}{}
    \Fn{\UpdateEncodeMap{E, domain}}{
        \ForEach{label in domain}{
            \If{label not in E}{
                $E[label] \leftarrow E.length$\;
            }
        }
        \KwRet{}\;
    }
    
    \SetKwProg{Fn}{Function}{:}{}
    \Fn{\EncodeLabels{RRs}}{
        Initialize $E$ to an empty hash map\;
        \ForEach{r in RRs}{
            \UpdateEncodeMap{E, r.rname}\;
            \If{$r.rtype = \bf{CNAME}$ \textbf{or} $r.rtype = \bf{DNAME}$}{
                \UpdateEncodeMap{E, r.rdata}
            }
        }
        \KwRet{E}\;
    }
\end{algorithm}

%具体而言，我们使用HashMap作为存储所有label编码的数据结构，将其命名为E。我们会尝试遍历所有所有可能的资源记录（7行），由于我们拥有所有需要验证的配置文件，获取到这些并不困难。随后对于每个资源记录<rname,rtype,rdata>，我们会分析rname中存在的所有label，对于那些未编码过的label，会赋予它新的值，即E的长度（9行，1-5行）。由于我们只会向E中添加新的label编码，而不会删除任何已经编码过的数据，所以每个label的编码必定不同。此外我们发现对于CNAME和DNAME记录，其rdata字段同样表示的是一个域名，因此我们对这类记录的rdata字段同样进行编码（10-11行）。
Specifically, we utilize a hashmap as the data structure to store the encoding of all labels, which we denote as $E$. We attempt to traverse all possible resource records within the zonefiles under verification (line 7). Subsequently, for each resource record\;$\langle \textit{rname}, \textit{rtype}, \textit{rdata}\rangle$, we analyze all the labels present in \textit{rname}. For those labels that have not been encoded, we assign them a new value, which corresponds to the current length of E (lines 9, 1-5). Since we only add new label encodings to E and never delete any already encoded data, each label's encoding is guaranteed to be unique. Furthermore, we observe that for CNAME and DNAME records, the \textit{rdata} field also represents a domain name. Therefore, we similarly encode the \textit{rdata} field of such records (lines 10-11).

% 最后，假设所有配置文件中，总共包含m个不同的label，且域名最长为n个label，那么我们只需要n * \lceil \log_2(\text{m}) \rceil + r 个变量就可以构建所有查询集合对应的BDD，其中r是编码rtype说所需的变量树，由于编码过程与编码label相似我们不做过多介绍。
The number of variables required to construct a BDD has a logarithmic relationship with the number of distinct labels. Assuming that all configuration files collectively contain $m$ distinct labels, and the longest domain name consists of $n$ labels, we only require $n * \lceil \log_2(m) \rceil + r$ variables to construct the BDD corresponding to all query sets. Here, $r$ represents the number of variables needed to encode $\textit{rtype}$. Since the encoding process for $\textit{rtype}$ is similar to that of labels, we will not elaborate further.
\\
\para{Details of storage LEC.}
% 在回答完如何将查询编码BBD后，我们介绍对于一个域名服务器是如何存储自身的LECs。首先我们需要对action做出一个正式的定义，来把相同aciton的查询进行聚合为一个LEC。定义1表示，一个action应该考虑其所在的域名服务器，zonefile，atype和adata。
We proceed to explain how a nameserver stores its own LECs. Initially, it is imperative to formally define an action to aggregate queries with identical actions into an LEC. Definition 1 stipulates that an action should consider the nameserver on which it resides, the zonefile, the $\textit{atype}$, and the $\textit{adata}$. 

\begin{defn}[Action]
We say $Action_{s,z}(q) = \langle \textit{atype},$ $\textit{adata}\rangle$ denotes the behavior generated by nameserver $s$ and its zonefile $z$ for $q$, where $\textit{atype} \in \{Answer, Delegate, RewriteC,$ $RewriteD, Refuse,$ $NonExist, ServiceFail,Loop\}$ donates the type of action (\eg, return the query answer Answer, delegate query to another nameserver or more nameservers Delegate, rewrite caused by CNAME record RewriteC, etc.) and $\textit{adata}$ stores the data needed for the action (\eg, IP that should be returned, the value after CNAME rewrite, etc.).
\end{defn}

%在拥有了对action定义后，我们设计了分层级的LEC存储结构。给定一个域名服务器ns，我们将其以<zones,refuse_rule>表示，其中zones是ns所有包含的所有zonefile对象，refuse_rule代表拒绝规则。对于具体的一个zonefile对象，其结构为<rules, hit, bdd, nx_rule>,其中rules表示该zonefile存储的所有的规则，hit为该zonefile在不受其他zonefile影响下能够处理的查询空间，bdd为该zonefile目前能够处理的查询空间，nx_rule对应为查询不存在的规则。对于一个具体的规则，其由三部分组成<hit, bdd, action>，其中hit与bdd的定义与zonefile中类似，而action表示改规则对应的行为。

With the definition of A\textsc{ction}, we design a hierarchical storage structure for LECs. Given a nameserver $ns$, we represent it as $\langle \textit{zones}, \textit{refuse\_rule}\rangle$, where $\textit{zones}$ encompass all the zonefile objects contained within $ns$, and $\textit{refuse\_rule}$ denotes the refusal rule. For a specific zonefile object, its structure is $\langle \textit{rules}, \textit{hit}, \textit{bdd}, \textit{nx\_rule}\rangle$, with $\textit{rules}$ representing all the rules stored in the zonefile, $\textit{hit}$ indicating the query space that the zonefile can handle independently of other zonefiles, $\textit{bdd}$ representing the current query space that the zonefile can process, and $\textit{nx\_rule}$ corresponding to the rule corresponds to rules for non-existent query results. A specific rule consists of three parts $\langle \textit{hit}, \textit{bdd}, \textit{action} \rangle$, where the definitions of $\textit{hit}$ and $\textit{bdd}$ are similar to those in the zonefile, and $\textit{action}$ signifies the behavior corresponding to that rule.

%通过这种分层式的架构使得我们能够高效的组织LEC。这也与我们action定义相符，即action与域名服务器和其上zonefile绑定，这也就意味着不同域名服务器上的本地等价类互不相同，相同域名服务器下的不同zonefile也拥有着不同的本地等价类。
This hierarchical architecture enables us to organize LECs efficiently. It also aligns with our definition of an action, which is tied to the nameservers and its zonefiles. This implies that LECs on different nameservers are distinct, and different zonefiles under the same nameserver also possess different LECs.

\subsection{LEC Construction}\label{subsec: LECConstruction}

% 为了正确的构造出每一个域名服务器，每一个zonefile对应的等价类，我们必须给出域名服务器是如何处理查询的逻辑。在RFC1034中已经给出了一个处理算法的推荐模板，然而具体的处理规则仍然是由各个操作系统和软件自行决定。本文将参考Groot给出的semantics，设计一套适用于LEC构建算法，并且保证完备性和互斥性，具体算法为算法2.
To construct the LECs for each nameserver and each zonefile, it is imperative to delineate the logic by which nameservers process queries. While RFC1034~\cite{rfc1034} provides a recommended template for a processing algorithm, the specific processing rules are ultimately determined by individual operating systems and software. This paper references the semantics provided by \groot{}~\cite{kakarla2020groot} to design a set of algorithms suitable for LEC construction, ensuring both completeness and mutual exclusivity. The specific algorithm is presented as Algorithm \ref{algleczone} and \ref{alglecns} (due to space constraints, we place Algorithm \ref{algsymbolic} and \ref{algleczone} in the appendix.).

% 在具体介绍这个算法之前，我们先解释一下其中提到的GetSpace(rname,types,flag)方法。该方法主要目的是将给定的域名和类型集合转为BDD表示，即一个查询集合。其中rname是以字符串形式存储的label序列，types为该查询集合中所有可能的类型。一个非常关键的参数flag代表是考虑子域以及如何考虑。以example.com为例，当flag=0时，代表的域名空间就是{example.com}本身；当flag=1时，代表域名空间为本身与子空间（即{example.com，*.example.com}）;当flag=2时，代表域名空间仅为其子空间，不包含本身（即{*.example.com}).
Before delving into the specifics of the algorithm, we first clarify the $GetSpace(\textit{rname}, \textit{types}, \textit{flag})$ method mentioned therein. The primary objective of this method is to convert a given domain name and set of types into a BDD representation, effectively a query set. Here, $rname$ is a sequence of labels stored as a string, and $types$ represents a set of query types included in the query set, where $\bf{all\_types}$ means all query types in the DNS system. A crucial parameter, $flag$, dictates whether and how subdomains are considered. Taking example.com as an instance, when $flag$=0, the domain name space represented is \{example.com.\} itself; when $flag$=1, the domain name space includes itself and its subdomains (i.e., \{\text{example.com.}, \text{*.example.com.}\}); and when $flag$=2, the domain name space is solely its subdomains, excluding itself (i.e., \{\text{*.example.com.}\}).

% 紧接着我们开始为域名服务器ns上的不同zonefile构造等价类。我们前面提到我们需要保证等价类之间的互斥性，因此在计算过程中我们维护一个变量remain_bdd用以保存剩余可用的域名空间。在整个过程中由origin（即zone负责的域）的长度来决定分配空间的先后顺序，符合最长前缀匹配。具体而言，每个空间先计算自身能够处理的最大空间zone_hit(行44），随后从remain_bdd中划走属于自己的空间，并更新remain_bdd（行45-46），这样较短的zonefile只能从剩余的空间中获取空间，即保证了两个zonefile的空间互不相交。请注意，zonefile能划得的空间与其包含的域名记录并无关系，也就是说ns为一个查询匹配zonefile时，不会考虑具体的资源记录。在为所有zonefile分配好查询空间后，剩余的空间的行为即为Refuse。
\begin{algorithm}
    \footnotesize
    \caption{Construct the LECs for a nameserver}\label{alglecns}
    \SetKwFunction{ConstructLECs}{ConstructLECs}
    \SetKwProg{Fn}{Function}{:}{}
    \Fn{\ConstructLECs{zonefiles}}{
        Sort $zonefiles$ by number of labels in origin\;
        $zones \leftarrow empty\ list$\;
        $remain\_bdd \leftarrow true$\;
        \ForEach{zonefile in zonefiles}{
            $zone\_hit \leftarrow $ \GetSpace{zonefile.origin, $\bf{all\_types}$, 1}\;
            $zone\_bdd \leftarrow zone\_hit\cap remain\_bdd $\;
            $remain\_bdd \leftarrow remain\_bdd-zone\_bdd$\;
            $rules, nx\_rule \leftarrow $ \text{$zonefile, zone\_bdd$}\;
            $zones.push(\langle rules,zone\_hit,zone\_bdd,nx\_rule \rangle)$\;
        }
        $refuse\_rule \leftarrow \langle remain\_bdd, remain\_bdd, \langle Refuse,()\rangle\rangle$\;
        \KwRet{$\langle zones, refuse\_rule\rangle$}\;
    }
\end{algorithm}
Next, in Algorithm \ref{alglecns}, we partition the space for the different zonefile objects on the nameserver $ns$. As mentioned earlier, we need to ensure mutual exclusivity among LECs. Therefore, during the computation process, we maintain a variable $remain\_bdd$ to keep track of the remaining available domain space. The order of space allocation is determined by the length of origin (i.e., the domain covered by the zone), following the principle of longest prefix matching. Specifically, each zone first calculates the maximum space it can handle, denoted as $zone\_hit$. Then, it subtracts its allocated space from $remain\_bdd$ and updates $remain\_bdd$ accordingly. This ensures that shorter zonefile objects can only acquire space from what remains, guaranteeing that the spaces of different zonefile objects are mutually exclusive. It is important to note that the space a zonefile can claim is unrelated to the domain name records it contains. In other words, when $ns$ matches a query to a zonefile, it does not take specific resource records into account. After allocating query spaces to all zonefile objects, the behavior for the remaining space is set to $Refuse$ .

% 然后我们考虑如何计算具体的每个LEC。给定一个zonefile以及他分得的查询空间，其本地等价类主要由其包含的资源记录进行分割。首先，我们将rname，rtype相同的记录进行聚合，因为显然这些记录属于同一个查询的结果（行11）。随后我们对每条聚合后的记录计算优先级（行14）。这部的目的与之前类似，同样是确定划分查询空间的顺序，以保证LEC之间互不相交。这里我们需要处理两种特殊情况DNAME和CNAME。对于DNAME记录，以<a.com.,DNAME,a.net.>为例，查询<r.a.com.,A>会被重写为<r.a.com.,A>，而查询<a.com.，DNAME>则直接返回结果a.net。因此DNAME实际上包含了两种可能的行为。同样的，对于CNAME也有类似的情况。所以我们会为这两种类型的记录生成两条规则，以两条记录不同rank呈现（行16-17）。最后我们开始具体计算每条规则，其过程与之前为各个zonefile划分空间相似（行22-37）。特别是由于我们之前计算rank时已经充分考虑将不同的action类型分配到不同的范围（行1-9），因此在计算规则时能够快速确认改规则对应的行为。至此，我们完成了对一个域名服务器构建LECs的完整过程。
Then, in Algorithm \ref{algleczone}, we consider how to compute each specific LEC. Given a zonefile and its allocated query space, its LECs are primarily divided based on the resource records it contains. First, we aggregate records with the same $rname$ and $rtype$, as these records evidently correspond to the result of the same query . Then, we calculate the priority (rank) for each aggregated record . The purpose of this step is similar to the earlier process: determining the order of query space division to ensure that LECs are mutually exclusive. Here, we need to handle two special cases: DNAME and CNAME. For DNAME records, for instance, $\langle \text{a.com.}, \text{DNAME}, \text{a.net.}\rangle$, a query such as $\langle \text{r.a.com.}, \text{A}\rangle$ will be rewritten to $\langle \text{r.a.net.}, \text{A}\rangle$, while a query $\langle \text{a.com.}, \text{DNAME}\rangle$ directly returns the result a.net. Thus, DNAME essentially involves two possible behaviors. Similarly, CNAME records exhibit comparable behavior. Therefore, we generate two rules for these types of records, with the two rules presented at different ranks. Finally, we compute each rule, following a process similar to that used for allocating space to zonefile objects. Notably, since the rank computation has already taken into account the assignment of different action types to separate ranges, the behavior corresponding to each rule can be quickly determined during rule computation. At this point, we complete the entire process of constructing LECs for a nameserver.

% 进一步的优化讨论。有时候，由于DNAME记录的存在，实际label的数量可能出现数量可能大于n。例如考虑记录<a.com., DNAME, a.a.com.>,这条记录会导致无限制的重写，即域名*.a.com会被改写成*.a.a.com，紧接着会继续被改写成*.a.a.com，不断循环，直到超过最大长度。可以发现，这个过程中域名的label数量不断增长。尽管我们可以在发现循环是及时终止，但是对于哪些跨越多个zonefile导致的循环我们难以直接发现，即我们无法确定在第一个循环结束时，查询域名中label数量的最大值。这时，我们将设置的权利交给用户，即用户可以设置冗余label数量rl（即域名长度最大为n+rl），来保证查询过程中不会出现因为label数量预留不足导致溢出。

\subsection{Further Optimization} 
\para{A more comprehensive label encoding scheme.} At times, due to the presence of DNAME records, the actual number of labels may exceed n. For instance, consider the record $\langle \text{a.com.}, $ $\text{DNAME}, \text{a.a.com.}\rangle$. This record can lead to unlimited rewriting, where the domain name *.a.com is rewritten as *.a.a.com, and then further rewritten as *.a.a.com, ad infinitum, until the maximum length is exceeded. It is evident that during this process, the number of labels in the domain name continuously increases. Although we can terminate the process upon detecting a cycle, it is challenging to directly identify cycles that span multiple zone files, meaning we cannot ascertain the maximum number of labels in the query domain name at the end of the first cycle. In such cases, we delegate the setting authority to the user, allowing them to specify the number of redundant labels $rl$ (i.e., the maximum domain name length is $n + rl$ labels), thereby ensuring that the query process does not overflow due to insufficient label count reservation.

%假设我们存在16个不同的label，最大label数量为4.那么最终在分配BDD变量时，x0-x3编码第1层label，x4-x7编码第二层label，以此类推。那么就引出一个问题，第一层作为顶级域名（例如com,net)通常偏少，而二三层子域的数量可能非常庞大，他们使用相同数量的变量是否造成浪费和额外开销？显然我们可以为不同层设置不同的label编码表格，来做到降低BDD变量数量的使用。例如第一层只有4种不同的label，第二层有8种不同的label，那么我们可以使用x0-x1来编码第一层的label，x2-x4编码第二层label。但是我们不得不考虑DNAME的情况，还是以<a.com., DNAME, a.a.com.>为例，b.a.com会被改写为b.a.a.com，也就是说虽然记录中的b出现在第三层，但是在实际过程中b是可以出现在第四层的，这时候就需要用户自行设置前几层使用独立的hashmap，而后续层共用一个hashmap，这个例子中从第三层开始都应该共用同一个hashmap。

Additionally, assuming we have 16 distinct labels with a maximum of 4 labels per domain, the BDD variables are allocated such that $x0-x3$ encode the first-level label, $x4-x7$ encode the second-level label, and so on. Since the number of labels at different levels may not be the same, we can reduce the number of BDD variables by employing separate label encoding tables for different levels. For instance, if there are only 4 distinct labels at the first level and 8 at the second level, we can use $x0-x1$ to encode the first-level labels and $x2-x4$ for the second-level labels. However, we must consider the implications of DNAME records. Taking $\langle \text{a.com.}, \text{DNAME}, \text{a.a.com.}\rangle$ as an example, b.a.com is rewritten as b.a.a.com. This means that although $b$ appears at the third level in the record, it can actually appear at the fourth level during the process. Therefore, users need to configure the system to use independent hashmaps for the initial levels and a shared hashmap for subsequent levels. In this example, starting from the third level, all subsequent levels should share the same hashmap.
\begin{figure}[t!]  % htbp 改为普通 figure 环境（去掉 *）
    \centering    
    % 单栏内显示子图
        \includegraphics[width=0.45\textwidth]{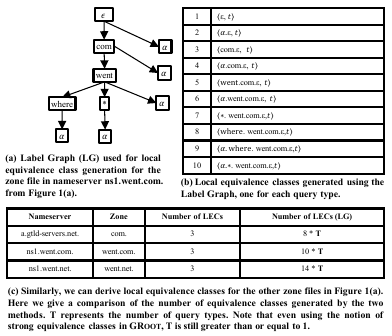}
        \label{fig: LECcomparision}  % 建议将子图标签放在 subfigure 内部
 %  \vspace{-10pt}
    \caption{Comparison of the LEC/EC in the example of Figure~\ref{fig: workflow} in detail. }  % 图注
    \label{fig: LECcomparision_main}  % 主图标签建议单独命名（避免与子图标签重复）
\end{figure}
\\
\para{Pursuing the ultimate number of LECs.}
% 无论是为了减少存储LECs的开销，还是降低遍历所有等价类的时间开销，我们都希望等价类的数量越少越好。为此我们可以基于3.2节中生成的LECs进行进一步的压缩操作。即我们最终可以将所有action相同的查询空间进行合并，进而做到最小等价类。由于BDD对逻辑操作的高效性，这个过程并不会花费特别多的时间。这时，对于域名服务器s生成的LEC集合 $C=\{c_1,c_2,...,c_n\}$满足以下约束：$1)\ \forall i,j:i \neq j \implies c_i\cap c_j=\emptyset$.$2)\ \forall c \in LECs: \forall q_1, q_2 \in c: q_1 \neq q_2 \implies Action_z (q_1 )=Action_z(q_2)$.
Whether to reduce the storage overhead of LECs or to lower the time complexity of traversing all equivalence classes, we aim to minimize the number of equivalence classes as much as possible. To achieve this, we can perform further compression on the LECs generated in Section \ref{subsec: LECConstruction}. Specifically, we can merge all query spaces with the same action into a single LEC, resulting in a minimal set of LECs. Thanks to the efficiency of BDD in performing logical operations, this merging process does not require significant computational effort. By comparing the EC generated by \groot{} in Figure~\ref{fig: LECcomparision_main} with the LEC generated by \system{} in Figure~\ref{fig: workflow_2}, it can be seen that \groot{} generates LECs for the 3 zonefiles that at least 2x more than \system{} generates. This shows that  \system{} has a significant effect on the compression of equivalence classes. At this point, the resulting LEC set for a nameserver $s$, denoted as $C=\{c_1,c_2,...,c_n\}$, satisfies the following constraints: 

$1)\ \forall i,j:i \neq j \implies c_i\cap c_j=\emptyset$. 

$2)\ \forall z \in s: \forall c \in C: \forall q_1, q_2 \in c: q_1 \neq q_2 \implies Action_{s,z} (q_1 )=Action_{s,z}(q_2)$. 

$3)\ \{q\ |\ q \in c \cap c \in C \}= The\ complete\ query\ space$. 

$4)\ \forall z \in s: \forall c_1,c_2 \in C: \forall q_1\in c_1,q_2\in c_2:c_1\neq c_2 \implies Action_{s,z}$ $(q_1) \neq Action_{s,z}(q_2)$. 

We call such a set the minimal LEC set.

% 1）表明LEC之间具有互斥性，这是由算法2保证的，我们在之前章节中以及提过。2）说明了LEC构造的正确性，即同一个LEC中的查询应该具有相同的行为。3）表明了其完备性，即所有LEC或者说所有查询子空间的并集为全集。4）通过将相同的action进行聚合，我们能保证任意两个LEC的行为并不相通。那么我们能够说，我们求得了最小LEC集合。
% 1) The mutual exclusivity among LECs is ensured by Algorithm 2, as discussed in previous chapters. 2) The correctness of LEC construction is affirmed, meaning that queries within the same LEC should exhibit identical behaviors. 3) The completeness is demonstrated, indicating that the union of all LECs, or all query subspaces, constitutes the entire set. 4) By aggregating identical actions, we guarantee that the behaviors of any two LECs are distinct. Therefore, we can assert that we have derived the minimal LEC set.

 %\vspace{-0.5cm}
\section{Symbolic Query Execution}\label{sec:Symbolic Query Execution}

% 给定一个包含多台nameservers的DNS子系统，我们在对每台nameserver构建完LECs后便应该开始寻找错误。我们引入了符号化执行的技术，以覆盖所有可能的查询，即做到全量覆盖，保证了验证的准确性。
Given a DNS subsystem comprising multiple nameservers, after constructing LECs for each nameserver, we should commence the search for errors. We introduce symbolic execution technology to cover all possible queries, achieving comprehensive coverage and ensuring the accuracy of verification.

\begin{algorithm}[t]
    \footnotesize
    \caption{\system{} symbolic execution logic}\label{algsymbolic}
    
    \SetKwFunction{Resolve}{Resolve}
    \SetKwProg{Fn}{Function}{:}{}
    \Fn{\Resolve{S, q, k}}{
        \ForEach{s $zones, refuse\_rule$ in S}{
            $q' \leftarrow q - refuse\_rule.bdd$\;
            \If{$q' \neq \emptyset$}{
                \Resolve{S, s, $q'$, k}\;
            }
            $q = q-q'$\;
        }
        \If{$q \neq \emptyset$}{
            $action \leftarrow \langle ServiceFail,()\rangle$\;
            Save relevant match information\;
        }
    }

    \SetKwProg{Fn}{Function}{:}{}
    \Fn{\Resolve{S, s, q, k}}{
        \If{s is not exist \textbf{or} $k<=0$} {
            $action \leftarrow \langle ServiceFail,()\rangle$\;
            Save relevant match information\;
            \KwRet{}\;
        }
        $visit \leftarrow \bigcup pq,$ where $pq$ represents all queries that have previously accessed $s$\;
        \If{$q \cap visit \neq \emptyset$}{
            $action \leftarrow \langle Loop,()\rangle$\;
            Save relevant match information\;
            \KwRet{}\;
        }
        \ForEach{rule $\langle hit,bdd,action\rangle$ in s}{
            $q' \leftarrow q \cap bdd$\;
            \If{$q' \neq \emptyset$}{
                Save relevant match information\;
                \If{action.atype = Delegate}{
                    \Resolve{S, rule.action.adata, $q'$, k-1}\;
                }
                \If{$action.atype \in \{RewriteC, RewriteD\}$}{
                    $q' \leftarrow $ Calculate new query after rewrite\;
                    \Resolve{S, $q'$, k-1}\;
                }
            }
        }
    }
    
\end{algorithm}

\subsection{A Symbolic Query Algorithm}
We now formally present how \system{} executes the symbolic query $q$ and returns the final result. As illustrated in Algorithm  \ref{alglecns}, the first function accepts a nameserver set $S$, a query $q$, and a fuel parameter $k$, which serves as a mechanism to mimic DNS behavior for guaranteeing query resolution termination. This function operates by processing the query $q$ across each nameserver $s \in S$ (lines 2-6). We will consider queries that cannot be resolved by the current DNS system and mark the behavior of these queries as $ServiceFail$ (lines 7-9). 

% 第二个函数回答了nameserver s如何处理查询q的过程。如果给定的s不存在DNS系统中，通常为将查询代理到一个空域名服务器，那么将终止查询（行11-14）。紧接着我们考虑查询出现循环的情况。在\system{}中每台nameserver会主动的保存其所有处理过的查询，同时在每次收到新查询时检测是否与之前的查询重复而造成循环。随后对于s持有的每条规则，我们将查看其是否与q匹配（行21-23）。当成功匹配到非终止节点时，查询过程将继续（行24-28）。例如如果查询q需要被代理到指定服务器，那么会根据action中存储的adata来获取一下跳的域名服务器，并进行递归解析。
The second function delineates the process by which the nameserver $s$ handles the query $q$. If the specified $s$ does not exist within the DNS system, typically implying the delegation of the query to a null nameserver, the query is terminated (lines 11-14). Next, we consider the scenario where query loops occur. In \system{}, each nameserver actively stores all the queries it has processed and checks for duplicates with each new query received to prevent loops (lines 15-19). Subsequently, for each rule held by $s$, we examine whether it matches $q$ (lines 21-23). Upon successfully matching a non-terminal node, the query process continues (lines 24-28). For instance, if the query $q$ necessitates delegating to a designated server, the next-hop nameserver is determined based on the $adata$ stored in the action, and recursive resolution is continued.

\subsection{Matching Optimization} \label{subsec: match opt}
%在符号化执行结束后，我们能够获取所有执行的traces。其中的每一条trace都表示着一个等价类q的行为，因此我们可以通过分析q的行为是否符合我们的断言来进行验证。
% \vspace{-2pt}
\begin{figure}[t]
    \centering
    \includegraphics[width=0.78\linewidth]{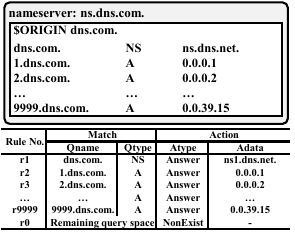}
    \caption{An example with a large number of LEC.}
    \label{fig:4.2.eg}
\end{figure}
% 尽管BDD在逻辑操作上的性能非常高效，但是当nameserver上规则的数量增长时，将查询与每条规则进行匹配仍然是一个非常耗时的操作。让我们考虑一个极端情况，见图3。对于域名服务器ns.extreme.org生成的LEC规则为r1,r2等等。域名服务器ns.extreme.com一共生成从rule0到rule9999一万条LEC。这时当ns.extreme.org收到符号化查询<*.extreme.org., A>后会将其改写为q': <alpha.extreme.com., A>并转发给namserver ns.extreme.com.紧接着，为了处理q'，我们需要执行1万次的逻辑与操作，每次的时间复杂度为O(|G|\dot |G|)，其中|G|为BDD中的节点个数。那么最终的时间复杂度将是10000 \dot O(|G|\dot |G|）。一旦该服务器需要处理多个查询，那么将会变成系统的一个性能瓶颈。
Although BDD is highly efficient in logical operations, as the number of rules on a nameserver increases, matching a query against each rule remains a time-consuming task. Let us consider an extreme case, illustrated in Figure \ref{fig:4.2.eg}. The nameserver ns.extreme.com generates a total of 10,000 LEC rules (r0 - r9999). When it receives a query $q:\langle \text{2.example.com.,A}\rangle$, to process $q$, we need to perform 10,000 logical AND operations, each with a time complexity of $O(|G| \cdot |G|)$, where $|G|$ is the number of nodes in the BDD. Thus, the final time complexity would be $10000 \cdot O(|G| \cdot |G|)$. If the server needs to handle multiple queries, this could become a performance bottleneck for the system.

%为此我们需要在进行正式匹配前进行有针对性的筛选，找出对q'可能存在影响的LEC。在Veriflow与Flash中都采用了一个multi-dimension prefix Trie的数据结构来在增量更新时高效发现筛选受影响的规则。我们借用其思想，我们发现有算法2生成的LEC均能够与一条具体的资源记录r相对应，因此我们可以通过r.rname筛选出与查询相关的LEC。具体来说，我们为查询q新增了prefix字段，表示该查询空间的公共前缀，初始为空，这样我们在进行匹配之前就能通过判断prefix是否与r.rname相匹配（即prefix是r.rname的子域或者r.rname是prefix的子域）来决定是否需要执行逻辑与操作。再次回到图3中的例子，查询<*.extreme.org., A, ''>在匹配到r2后将生成q':<alpha.extreme.com., A,alpha.extreme.com.>。这里我们定义，当查询匹配到Delegate action后，prefix将被设置为r.rname；当查询匹配到RewriteC或者RewriteD类型的action后，prefix将被设置为r.radata；其余类型的action不会改变prefix。当q'再次到达ns.extreme.com时，我们会执行一万次的比较操作，最终只会与rule2进行逻辑与操作。这样我们的时间复杂度将将降低为O(10000*n + |G| *|G|),其中n为rname的长度。

To address this, we need to perform targeted filtering before formal matching to identify LECs that may potentially affect $q$. Both Veriflow~\cite{veriflow} and Flash~\cite{flash} utilize multi-dimension prefix Trie to efficiently discover or filter rules. Borrowing from this concept, we observe that LECs generated by Section \ref{subsec: LECConstruction} can correspond to a specific record $r$, allowing us to filter LECs related to the query through $r.rname$. Specifically, we introduce a $prefix$ field to the query $q$, which denotes the common prefix of the query space and is initially empty. This allows us to determine whether to perform the logical AND operation by assessing if the $prefix$ matches $r.rname$ (i.e., the $prefix$ is a subdomain of $r.rname$ or $r.rname$ is a subdomain of the $prefix$) before proceeding with the matching process. 
Returning to the example in Figure \ref{fig:4.2.eg}, We assume that the query \( q: \langle \text{1.extreme.com.}, A, \text{1.extreme.com.} \rangle \) is derived from the query \( \langle \text{*.extreme.net.}, A \rangle \) through rewriting due to the record \( \langle \) \( \text{*.extreme.net.}, \text{CNAME}, \text{1.extreme.com.} \rangle \), meaning it matches a \( RewriteC \) action. Here, we define that when a query matches a $Delegate$ type action, the prefix is set to $r.rname$; when it matches a $RewriteC$ or $RewriteD$ type action, the prefix is set to $r.radata$; other types of actions do not alter the prefix. When $q$ reaches ns.extreme.com again, we perform 10,000 comparison operations, ultimately only executing the logical AND operation with r2. This reduces our time complexity to $O(10000\cdot n + |G| \cdot |G|)$, where $n$ is the length of $rname$.

\subsection{Property Check}  \label{subsec: pc}

Upon completing symbolic execution, we obtain all execution traces, each representing the behavior of an EC \( q \). These traces enable verification by analyzing whether \( q \)'s behavior adheres to predefined assertions. Specifically, each trace captures the sequence of query resolution steps, including delegation, rewriting, and response generation, which can be systematically compared against expected outcomes. Finally, we list in Table \ref{table:error_example} a few typical error representations and how they are verified in our model.

\begin{table*}[htbp]
\footnotesize
\begin{tabular}{@{}lp{13cm}@{}}
\toprule
\textbf{Error} & \textbf{Description} \\
\midrule
% \textbf{Cyclic Zone Dependency} & For a trace that there exists two logs that are the same.\\
\textbf{Delegation Inconsistency} &
For a trace that the parent node with the answer type is Ref and the child with the answer type is ANS but do not have the same set of NS and A records for delegation.\\
\textbf{Lame Delegation} &
For a trace that the answer type returns REFUSED.\\
\textbf{Missing Glue Records} &
For a trace that the answer type in any log is NS but not followed by the A (glue) records.\\
\textbf{Non-Existent Domain} &
For a trace that the answer returns NXDOMAIN.\\
\textbf{Cyclic Zone Dependency} &
For a trace that there exists two logs that are the same.\\
\textbf{Rewriting Loop} &
For a trace that there exists a rewrite that result in a loop.\\
\textbf{Query Exceeds Maximum Length} &
For a trace that the length of the input query or the number of the labels exceeds the maximum number. \\
% \textbf{Zero Time To Live} &
% For a trace that return the answer with TTL value is 0.\\
\textbf{Rewrite Blackholing} &
For a trace after the rewrite log the next log returns NXDOMAIN.\\

\bottomrule
\end{tabular}
\caption{Types of DNS misconfiguration that our framework support. We conclude from \groot{}~\cite{kakarla2020groot}, the formal method~\cite{liu2023formal} and the RFC~\cite{rfc8484}. }
\centering
\label{table:error_example}
\end{table*}
\section{Incremental Verification}\label{sec:Incremental Verification}
% 我们提出了一种高效算法，来在配置文件发生更新时进行快速增量验证。该算法的主要挑战为：1）如何确定配置文件在增加或者删除资源记录后会影响原有的哪些记录？2）如何避免重新进行全量符号化执行？
We propose an efficient algorithm for fast incremental verification when configuration files are updated.

\subsection{LECs Update}
% 我们目前主要考虑的更新情况为对zonefile进行增加或者删除资源记录的情况。本小节暂时不涉及对nameserver中的zonefile进行整体删除和增加。
Currently, our primary focus is on updates involving the addition or deletion of resource records within a zonefile. This section does not address the wholesale deletion or addition of zonefiles within a nameserver.

% 基本思想：基于优先级的抢夺式更新。在算法2中，我们能够推导出对于每条atype为非NonExist的规则r有以下两个性质：1）每条rule的r.hit是相互独立的；2）每条rule暗含一个rank字段，其影响着r.bdd的计算，即rank较高的rule优先计算。我们将atype为NonExist的规则记为nr，记其优先级为nr.rank = -1.
\para{Basic Idea: Priority-based preemptive update.} In Algorithm 2, we can deduce two properties for each rule $r$ where $atype$ is not $NonExist$: 1) The $r.hit$ of each rule is mutually independent; 2) Each rule implicitly contains a field $rank$, which influences the calculation of $r.bdd$, meaning that rules with higher ranks are computed first. We denote rules with $atype$ as $NonExist$ as $nr$. We denote the rule with type $NonExist$ as $nr$, and assign its priority as $nr.rank = -1$.

% 当我们像zonefile z中插入一条资源记录<rname1, rtype1, rdata1>时，如果z中已经存在另外一条记录<rname2, rtype2, rdata2>，满足 rname1 = rname2 and rtype1 = rtype2; 或者当删除<rname1, rtype1, rdata1>时，满足 rname1 = rname2 and rtype1 = rtype2 and adata1 != adata2，LECs不变，仅仅更新对应规则的adata字段。否则我们会向z中加入或者删除规则r，等式1给出了如何计算r.bdd。当删除规则时，由于规则已经事先计算好，并且存在z中，因此我们能够直接获取r.bdd。对于增加规则的情况，我们会从nr和其他优先级较低的规则r'中抢夺查询空间。
\para{Step 1. Compute the query space for the update rule.} When we insert a resource record $\langle rname1, rtype1, rdata1 \rangle$  into zonefile $z$, if $z$ already contains another record $\langle rname2, \\ rtype2, rdata2 \rangle$  that satisfies $rname1 = rname2 \wedge rtype1 = rtype2$; or when deleting $\langle rname1, rtype1, rdata1 \rangle$, if it satisfies $rname1 = rname2 \wedge rtype1 = rtype2 \wedge adata1 \neq adata2,$ the LECs remain unchanged, and only the $adata$ field of the corresponding rule is updated. Otherwise, we add or delete a rule $r$ from $z$, and Equation (\ref{eq: incbdd1}) provides the method for calculating $r.bdd$. When deleting a rule, since the rule has been precomputed and exists within $z$, we can directly retrieve $r.bdd$. For the case of adding a rule, we will preempt the query space from $nr$ and other lower-rank rules $r'$.
%\vspace{-0.1cm}
\begin{equation}\label{eq: incbdd1}
    r.bdd = 
    \begin{cases}
        r.bdd, & \text{delete rule } r\\\\
        r.hit \wedge \left( \bigvee\limits_{r.rank > r'.rank} r'.bdd \right), & \text{add rule } r
    \end{cases}
\end{equation}

\para{Step 2. Update of affected rules.}
% 由于我们在计算r.bdd时对rank较低的规则所拥有的空间进行了抢占操作，因此我们现在需要将这些受到空间抢占的规则的bdd进行更新。等式2给出了这部操作的算法。在删除规则r后，r所持有的空间将被释放，这时所有rank低的规则r'都会按顺序抢占被释放出的查询空间。也就是说，r'不能获取优先级更高的规则r''所想要抢占的空间r''.hit。对于添加规则的情况，r'从自身删除被r夺取的查询空间即可。
Since we performed space preemption on the rules with lower ranks while calculating \( r.bdd \), we now need to update the BDDs of those rules that were subject to space preemption. Equation (\ref{eq: incbdd2}) provides the algorithm for this operation. After the deletion of rule \( r \), the space held by \( r \) will be released, and all rules with lower ranks (\( r' \)) will sequentially preempt the released query space. In other words, \( r' \) cannot preempt the space \( r''.hit \) that is intended to be preempted by a higher-priority rule \( r'' \). For the case of adding a rule, \( r' \) simply removes the query space previously taken by \( r \).
%\vspace{-0.15cm}
\begin{equation}\label{eq: incbdd2}
  r'.{bdd} = 
  \begin{cases}
    \begin{split}
      r'.{bdd} \vee \Bigl( r'.{hit} \wedge r.{bdd} \wedge {} \\ 
      \neg \Bigl(\bigvee\limits_{r''.{rank} > r'.{rank}} r''.{hit} \Bigr)\Bigr),
      \end{split} & \text{delete rule } r \\
    r'.{hit} \wedge \neg r.{bdd}, & \text{add rule } r
  \end{cases}
\end{equation}

% 优化计算流程。正如我们在4.2节中说的，逻辑操作仍然需要一个不小的时间开销，因此接下来我们将介绍一些tricky的方法来减少逻辑操作量。我们观察等式3中计算所有r'.bdd析取的过程。我们可以将整个过程转变为类似为r.hit and r1'.bdd and r2'.bdd ...的形式。这样我们就可以在每次合取操作后判断r.bdd是否等于r.hit来决定是否提前结束，以降低计算的复杂度。对于等式4，我们可以以类似的方法，追踪r释放或者夺取的空间（初始为r.bdd)，将其记为s。在每次对r'.bdd完成更新后，同样更新s，当s为空时便可以终止对r'的遍历。而对于r''的析取，我们可以采用动态编程的思想，来防止每次对r'的更新都需要重新计算整个析取。此外，还可以通过比较规则之间的rname字段来筛选有关的/受影响的规则来加速计算。
\para{Optimizing the computation process.} As mentioned in Section \ref{subsec: match opt}, logical operations still incur a significant time overhead. Therefore, we will now introduce some tricky methods to reduce the amount of logical operations. We begin by analyzing the process of computing the disjunction of all \( r'.bdd \) terms as shown in Equation (\ref{eq: incbdd1}). We can transform the entire process into a form similar to "$ r.hit \wedge r_1'.bdd \wedge r_2'.bdd \ ... $". By doing so, after each conjunction operation, we can check if \( r.bdd \) is equal to \( r.hit \) in order to decide whether to terminate early, thus reducing the computational complexity. For Equation (\ref{eq: incbdd2}), we can adopt a similar approach by tracking the space released or preempted by \( r \) (initially \( r.bdd \)), which we denote as \( s \). After each update to \( r'.bdd \), we update \( s \) accordingly. When \( s \) becomes empty, we can terminate the traversal of \( r' \). As for the disjunction of \( r'' \), we can apply dynamic programming techniques to avoid recalculating the entire disjunction each time \( r' \) is updated. Additionally, we can accelerate the computation by comparing the \( rname \) fields between rules to filter out unrelated or unaffected rules.

% zonefile级别的增加与删除。接下来我们将讨论如何将上述方法扩展到添加或删除一个nameserver ns上的zonefile。我们在算法2中对ns上的zonefiles按照origin长度进行排序，也就说我们可以将origin所在DNS系统中的层级认为是优先级。例如我们可以记顶级域名（com）的zonefile的rank为1，二级域名（went.com)的zonefile的rank为2。这样zonefile也同时拥有hit,bdd,rank三个属性，那么我们可以直接套用等式1和2进行LEC的增量更新。对于bdd更新过的zonefile z，我们则可以认为向z中隐式添加一条rank为512的规则（添加的规则对应的bdd为z被抢夺的查询空间）或者隐式删除一条rank为0的规则（删除的规则对应的bdd为z新获取的查询空间）。
\para{Addition and deletion at the zonefile level.} To extend the aforementioned methodology to the addition or deletion of zonefiles on a nameserver $ns$, we discuss zonefile-level operations. In Algorithm \ref{alglecns}, zonefiles on $ns$ are sorted by the hierarchical depth of their origins within the DNS system, where the origin's hierarchy determines its rank. For instance, a top-level domain (\eg, com) is assigned a rank=1, while a second-level domain (\eg, went.com) is assigned rank=2. Each zonefile is endowed with three attributes: $hit$, $bdd$, and $rank$, enabling direct application of Equations (\ref{eq: incbdd1}) and (\ref{eq: incbdd2}) for incremental updates of LECs. For a zonefile $z$ whose $bdd$ is updated, this can be interpreted as either implicitly adding a rule with $rank=512$ (where the rule’s $bdd$ corresponds to the preempted query space from $z$) or implicitly deleting a rule with $rank=0$ (where the rule’s $bdd$ corresponds to the newly acquired query space for $z$). Next, just update all the rules in $z$ in the same way you would add or delete a rule.
\subsection{Incremental Execution} %Verification
% 为了避免进行全量的符号化执行流程，我们需要仅考虑哪些由于LECs更新所影响的trace。对于更新的zonefile z，我们会遍历所有与其相关的log L，并获取L对应的输入查询q。我们将L及其之前的logs按照顺序组成一个子trace T_prev，并删除所有原先包含L的traces。随后我们以z所在nameserver开端，q为输入，重新执行并生成后续的logs。记后续生成的logs组成的trace为T_next,那么最终新生成的trace即为T= T_prev + T_next. 至于验证过程与之前的差异不大。
% To avoid performing a full symbolic execution, we need to focus only on the traces affected by updates to the LECs. For an updated zone file \( z \), we traverse all the logs \( L \) related to it and retrieve the corresponding input query \( q \). We then assemble \( L \) along with its preceding logs into a sub-trace \( T_{\text{prev}} \), and remove all traces that originally contain \( L \). Next, starting from the nameserver where \( z \) is located and using \( q \) as the input, we re-execute the process and generate the subsequent logs. Let the trace composed of the newly generated logs be \( T_{\text{next}} \). Therefore, the final new trace is given by \( T = T_{\text{prev}} + T_{\text{next}} \). As for the verification process, it does not differ significantly from the previous approach.
To enhance verification efficiency while circumventing the computational burden of full symbolic execution, we propose a delta-update mechanism that selectively regenerates execution traces impacted by modifications to LECs.

\para{Phase 1: Affected trace identification.} When a zone file \( z \) is modified, the system analyzes dependencies via log relationships. It retrieves historical logs \( L \) linked to \( z \) and extracts the maximal preceding sub-trace \( T_{\text{prev}} \), capturing nameserver states and query history. Input queries \( Q = \{q_1, q_2, ..., q_n\} \) are filtered using BDD-based checks to isolate relevant queries.

\para{Phase 2: Partial symbolic re-execution.} Recomputation begins at the nameserver hosting \( z \), with execution parameters restored from \( T_{\text{prev}} \). Input queries \( Q \) are replayed through the updated LEC configuration, pruning unchanged BDD subspaces. The system generates version-tagged logs \( T_{\text{next}} \).

\para{Phase 3: Trace integration and verification.} The final synthesis aligns \( T_{\text{prev}} \) and \( T_{\text{next}} \) to reconstruct event timelines. Verification follows the approach outlined in section \ref{subsec: pc}.

\begin{figure*}[htp]
    \centering
    % 第一个子图组，共享标题
    \begin{minipage}{0.48\textwidth}
        \centering
        \begin{subfigure}[b]{1\textwidth}
        % \vspace{0.4cm}
            \includegraphics[width=\textwidth]
            {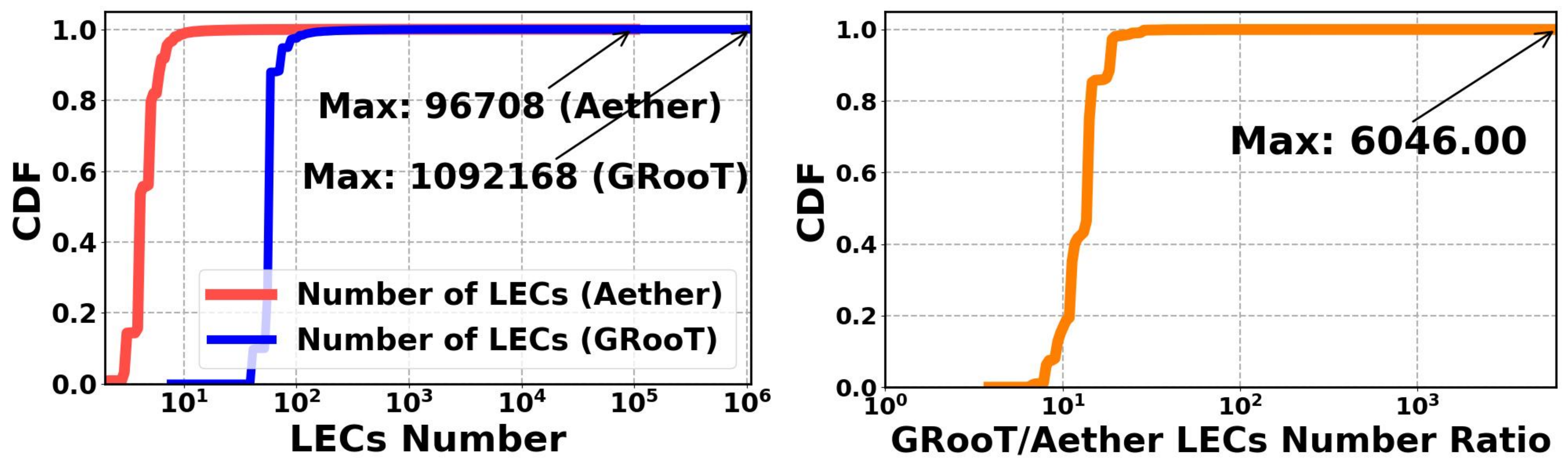}
             \caption{Burst update LEC/EC number and ratio.}
             \label{fig:burst_lecnum}
        \end{subfigure}
        % \begin{subfigure}[b]{0.45\textwidth}
        %     \includegraphics[width=\textwidth]{figures/temporary/burst/cttime_ration.png}
        %      \caption{Burst update LEC construction time ratio.}
        % \end{subfigure}
         %\caption{ }
    \end{minipage}
    % \hfill
 % \hspace{0.00000000000000000000000000001\textwidth}
    % 第二个子图组，共享标题
    \begin{minipage}{0.48\textwidth}
        \centering
        \begin{subfigure}[b]{1\textwidth}
            \includegraphics[width=\textwidth]{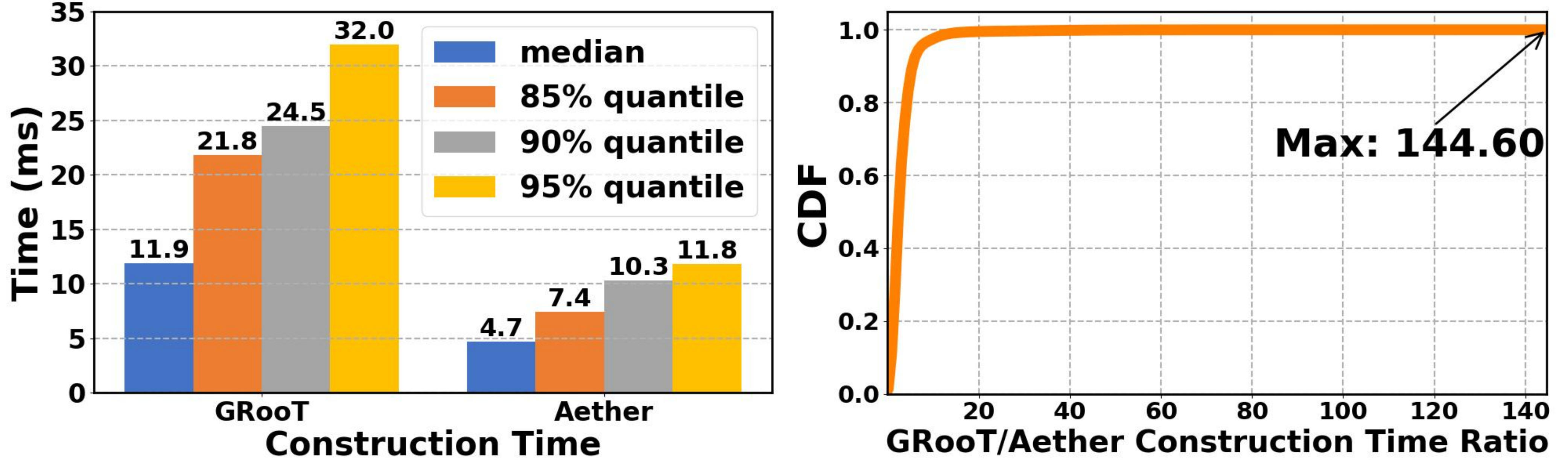}
             \caption{Burst update LEC/EC construction time and ratio.}
              \label{fig:burst_ct}
        \end{subfigure}
        % \begin{subfigure}[b]{0.45\textwidth}
        %     \includegraphics[width=\textwidth]{figures/test_10_8_2.pdf}
        %     % \caption{4}
        % \end{subfigure}
    \end{minipage}
        \begin{minipage}{0.48\textwidth}
        \centering
        \begin{subfigure}[b]{1\textwidth}
            \includegraphics[width=\textwidth]{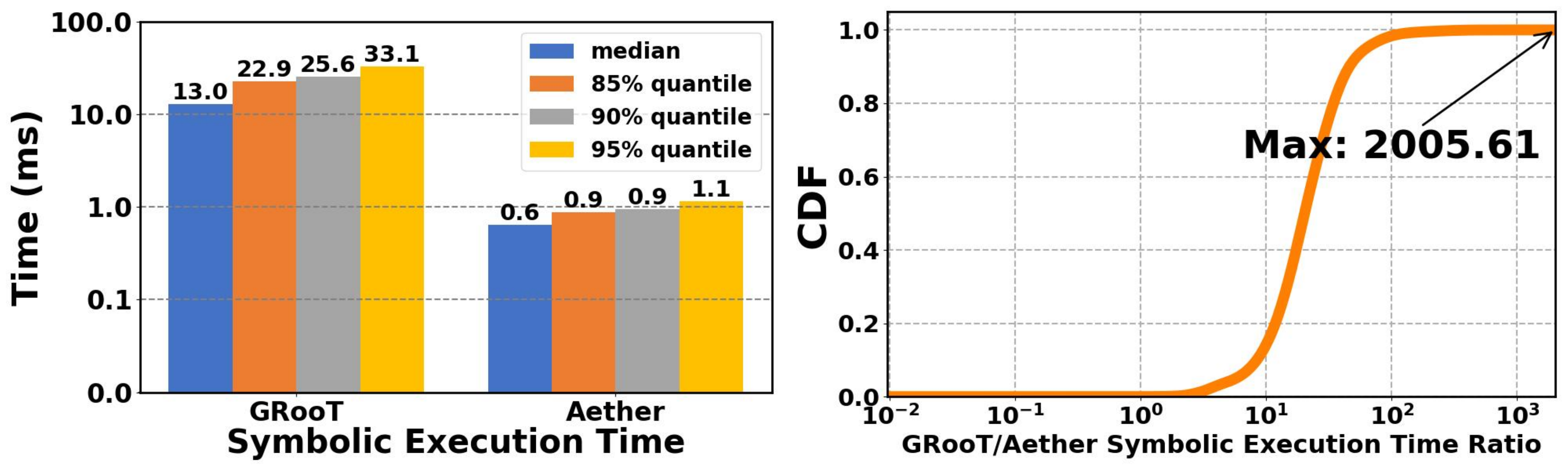}
             \caption{Burst update symbolic execution time and ratio.}
             \label{fig:burst_se}
        \end{subfigure}
        % \begin{subfigure}[b]{0.45\textwidth}
        %     \includegraphics[width=\textwidth]{figures/test_10_8_2.pdf}
        %     % \caption{4}
        % \end{subfigure}
    \end{minipage}    \begin{minipage}{0.48\textwidth}
        \centering
        \begin{subfigure}[b]{1\textwidth}
            \includegraphics[width=\textwidth]{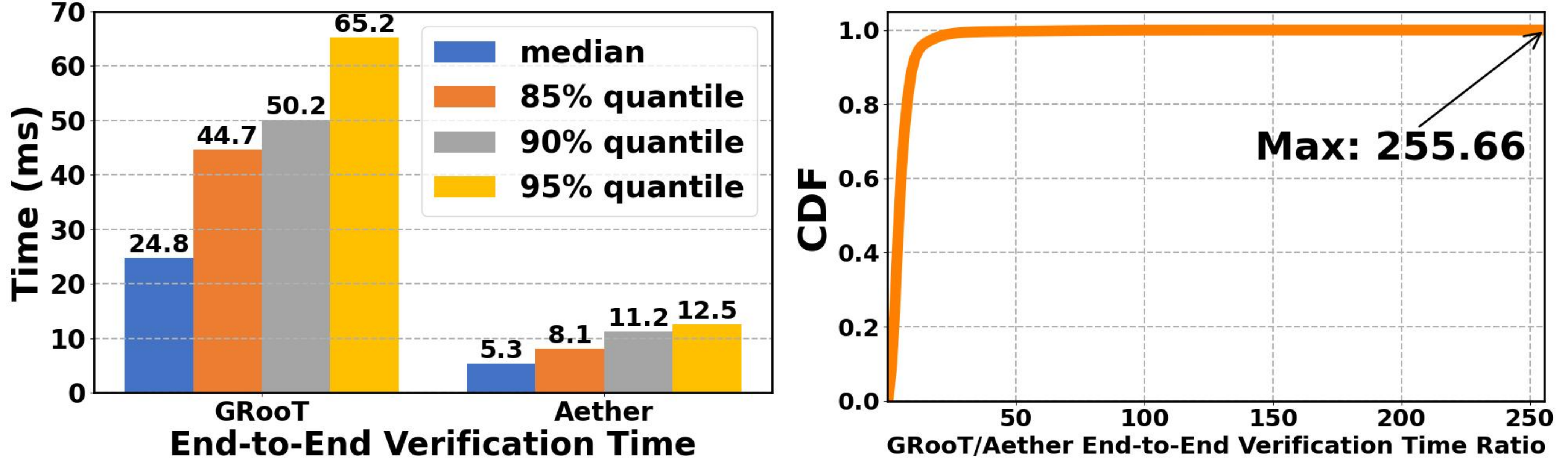}
             \caption{Burst update end-to-end verification time and ratio.}
             \label{fig:burst_e2e}
        \end{subfigure}
        % \begin{subfigure}[b]{0.45\textwidth}
        %     \includegraphics[width=\textwidth]{figures/test_10_8_2.pdf}
        %     % \caption{4}
        % \end{subfigure}
    \end{minipage}
    \vspace{-5pt}
     \caption{Burst update experiment result using Census dataset.}
\end{figure*}
\section{Evaluation}
%We conduct extensive evaluations on \system{}. Specifically, we study four questions: (1) What is the capability of \system{} in verifying generic invariants? (\S\ref{sec: demonstrations}) (2) What is the performance of \system{} in a testbed with different types of network devices mimicking a real-world network? (\S\ref{sec: experiments}) (3) What is the performance of \system{} in various real-world, large networks? (\S\ref{sec: simulations}) (4) What is the overhead of running \system{} on network devices? (\S\ref{sec: microbenchmarks})
 We implement a prototype of \system{} using Rust based on the OxiDD~\cite{oxidd24} framework and conduct extensive evaluations on \system{}. For all the zones in the dataset, the initial input query is $q:\langle *，* \rangle$, which means that for all possible domains and all the types of query, we want to know their response. Specifically, we study the following questions: (1) What is the capability of \system{} in verifying DNS configuration files? (\S\ref{subsec:Capability Experiment}) (2) What is the performance of \system{} in a real-world DNS dataset? (\S\ref{subsec:Performance Experiment}) 
% ================= 段落1 ==================
%\subsection{Functionality Demonstrations}
%\label{sec: Functionality Demonstrations}
%We build a network of 5 nameservers in Figure \ref{}. We run demos to verify $(1)$ missing glue, delegation inconsistency from $A$ to $B$, $(2)$ cyclic zone Dependency, the repeated query from $C$ to $A$, $(3)$ rewrite loop, rewrite blackholing, the repeated query from C to D, $(4)$ domain overflow, domain overflow at nameserver at every nameserver, $(5)$ answer inconsistency, inconsistent RRsets from $B$ to $D$ and $E$. We run each demo with the correct and erroneous dataset. The \system{} always verifies the misconfiguration bug.

% ================= 段落2 ==================
%\subsection{Verification Experiments}
%\label{subsec: Verification Experiments}
%We use 3 servers to build the distributed DNS configuration verification experiment testbed (Server1-Server3) and the remaining one as the centralized DNS misconfiguration experiment testbed (Server4). We also add the propagation latencies for the dataset based on its root DNS server location. We verified the domain overflow at the nameserver and repeated the query.

% To keep the experiment environment the same, we chose two servers with the same configuration to perform the DNS configuration verification experiments of our prototype and the verification experiments of \groot{}.

\subsection{Capability Evaluation} \label{subsec:Capability Experiment}
To maintain the uniformity of the experimental environment, the verification experiments all run on the platform configured with Ubuntu 20.04.3 LTS (kernel version 5.13.0-41-generic), equipped with two Intel Xeon Gold 6126 CPU (2.60GHz) and 128GB of DDR4 memory.
% \vspace{-5pt}
\begin{table}[htbp]
\small
\begin{tabular}{@{}p{0.23\textwidth} p{0.1\textwidth} p{0.1\textwidth}@{}}
\toprule
\textbf{Error} & \textbf{\groot{}} & \textbf{\system{}} \\
\midrule
% \textbf {Zero Time To Live} & {\textcolor{red}{0}}$^{\textcolor{red}{\star}}$\\
% \textbf{Cyclic Zone Dependency} & For a trace that there exists two logs that are the same.\\
\textbf{Delegation Inconsistency} &
{\textcolor{red}{$\times$}}$^{\textcolor{red}{\star}}$ & {\textcolor{red}{$\checkmark$}}$^{\textcolor{red}{\star}}$\\
\textbf{Lame Delegation} &
{\textcolor{red}{$\checkmark$}}$^{\textcolor{red}{\star}}$ & {\textcolor{red}{$\checkmark$}}$^{\textcolor{red}{\star}}$ \\
% \textbf{Non-Existent Domain} &
% {\textcolor{red}{0}}$^{\textcolor{red}{\star}}$\\
% \textbf{Cyclic Zone Dependency} &
% {\textcolor{red}{0}}$^{\textcolor{red}{\star}}$\\
\textbf{Rewriting Loop} &
{\textcolor{red}{$\checkmark$}}$^{\textcolor{red}{\star}}$& {\textcolor{red}{$\checkmark$}}$^{\textcolor{red}{\star}}$\\
% \textbf{Query Exceeds Maximum Length} &
% {\textcolor{red}{0}}$^{\textcolor{red}{\star}}$ \\
% \textbf{Answer Inconsistency} &
% {\textcolor{red}{0}}$^{\textcolor{red}{\star}}$\\
% \textbf{Zero Time To Live} &
% For a trace that return the answer with TTL value is 0.\\
\textbf{Rewrite Blackholing} &
{\textcolor{red}{$\checkmark$}}$^{\textcolor{red}{\star}}$& {\textcolor{red}{$\checkmark$}}$^{\textcolor{red}{\star}}$\\
\hdashline
\textbf{Number of rewrites >=2} &
{\textcolor{orange}{$\checkmark$}}$^{\textcolor{orange}{\dagger}}$ &
{\textcolor{orange}{$\checkmark$}}$^{\textcolor{orange}{\dagger}}$\\
\textbf{Number of hops >=2} &
{\textcolor{orange}{$\checkmark$}}$^{\textcolor{orange}{\dagger}}$ &
{\textcolor{orange}{$\checkmark$}}$^{\textcolor{orange}{\dagger}}$\\

\bottomrule
\end{tabular}
\caption{Errors checked on the campus dataset and the number of cases \system{} reported. Cases in red with marker $^\star$ are errors while orange with marker $^\dagger$ are warnings.}
\label{table:capability}
\centering
\label{additional_experiment}
% \vspace{-5pt}
\end{table}

To evaluate the capability of \system{}, we set up the experiment on the university dataset and evaluate the property that \system{} verified.\\
\para{Dataset.} We collect the DNS information from a university with more zonefiles from multiple nameservers and any wildcard records to show the capability of \system{}. The university dataset obtained from the laboratory features multiple levels of subdomains (more than 2). It includes 9,700 zonefiles with 134,349 resource records in total.\\
\para{Metric.} We record the DNS configuration errors that \system{} and \groot{} verified in this dataset.\\
\para{Experiment result.} The results in Table \ref{table:capability} demonstrate that \system{} successfully verified 4 types of DNS configuration errors and 2 types of warnings, while \groot{} verified 3 types of DNS configuration errors and 2 warnings. Through analyzing Groot's code and design, we discovered discrepancies between the implementation and the documentation. This led to an incomplete detection of the Delegation Inconsistency error. The system only checks whether the parent and child domains are about the same domain name by verifying the consistency of the first record's name. Consequently, if the child domain's zone file is located in a subsequent position, it may be missed during verification.\\
\para{Additional Experiment.}\label{subsec:Additional Experiment} As mentioned in the previous section, the definition of \groot{} is unsound. To demonstrate this, we have set up an additional experiment. The example illustrated in our workflow (Figure \ref{fig: workflow}) serves as a representative case. Through this experimental result, we intend to illustrate not only the unsoundness of \groot{} but also the necessity of introducing the concept of local equivalence classes. %\\

%\para{Dataset.} 
We transform the examples in the workflow into a dataset containing three zonefiles.%\\
%\para{Metric.} 
we run the source code of \groot{} and \system{} to get the verification result.%\\
%\para{Experiment result.} 
We check the default properties of \groot{}, as it claims that the default properties can detect all loops. However, during the execution, we find that \groot{} did not verify the existence of a rewriting loop. Then we check for the "Number of Hop" and the result shows \groot{} found 3 number of hop errors. In \system, we identify 6 errors, one of which is rewrite blackholing, another is a rewrite loop, and the remaining 4 are the number of hop errors. This indicates that, under the local equivalence classes we defined, \system is capable of verifying these errors.
%\todo{
%We run demos to verify the common errors in Table~\ref{table:error_example} that we have concluded from \groot{}~\cite{kakarla2020groot}, the formal method~\cite{liu2023formal} and the RFC~\cite{rfc8484}. 
%$(1)$ delegation consistency, the parent and the child files do not have the same set of NS and A records for delegation, $(2)$ lame delegation, the name server that is authoritative for a zone does not provide authoritative answers $(3)$ rewrite loop, a query that is rewritten in a loop, $(4)$ rewrite blackholing, a query that is eventually rewritten and does not exist and DNS returns NXDOMAIN.
%We run the \system{} in the real-world dataset. \system{} can successfully catch some errors.
%}

\subsection{Performance Evaluation} \label{subsec:Performance Experiment}
To evaluate the performance of \system{}, we set up the experiment on the open-sourced dataset and evaluate the end-to-end verification time in the scenario of burst update and incremental update.\\
\para{Dataset.} We get the dataset from the real world named Census~\cite{dnscensus2013}. The census consists of 1,368,523 zones and over 65 million resource records. However, according to our test results with Bind9~\cite{bobcares2020}, some files in this dataset violate the named-checkzone directive (\eg, quotes are not enclosed). These zones are defined as invalid zones, and we filter them out in our experiments (remaining 1,186,517 zones) because they are syntactically incorrect or incomplete. Thees files cannot be parsed successfully.
% This means that for some zones, the syntax and completeness of their configuration files are problematic .Census was collected from real-time DNS queries during 2012-2013, filtered by timestamps, and processed by DNS name space hierarchy. The dataset consists of 1,368,523 zones and over 65 million resource records. 
\\
\para{Metric.} We evaluate the LEC construction time, symbolic execution time, and end-to-end verification time (LEC construction time + symbolic execution time) in burst and incremental update scenarios to represent the overall performance. We also compare the number of LEC of \system{} with the number of EC of \groot{} to show our contribution to the definition of LEC.\\

\begin{figure*}[htp]
    \centering
    % 第一个子图组，共享标题

        \begin{minipage}{0.7\textwidth}
        \centering
        \begin{subfigure}[b]{1\textwidth}
            \includegraphics[width=\textwidth]{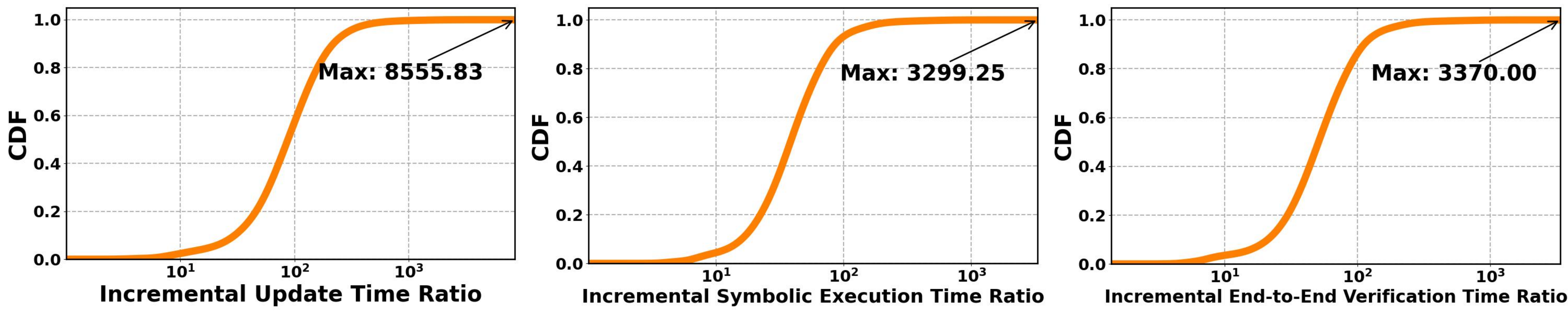}
             \caption{Incremental update construction time ratio, symbolic execution ratio and end-to-end verification time ratio. All the ratios are calculated as: \groot{}/\system{}.}
             \label{fig:incre_combin}
        \end{subfigure}
        % \begin{subfigure}[b]{0.45\textwidth}
        %     \includegraphics[width=\textwidth]{figures/test_10_8_2.pdf}
        %     % \caption{4}
        % \end{subfigure}
    \end{minipage}    \begin{minipage}{0.24\textwidth}
        \centering
        \begin{subfigure}[b]{1\textwidth}
            \includegraphics[width=\textwidth]{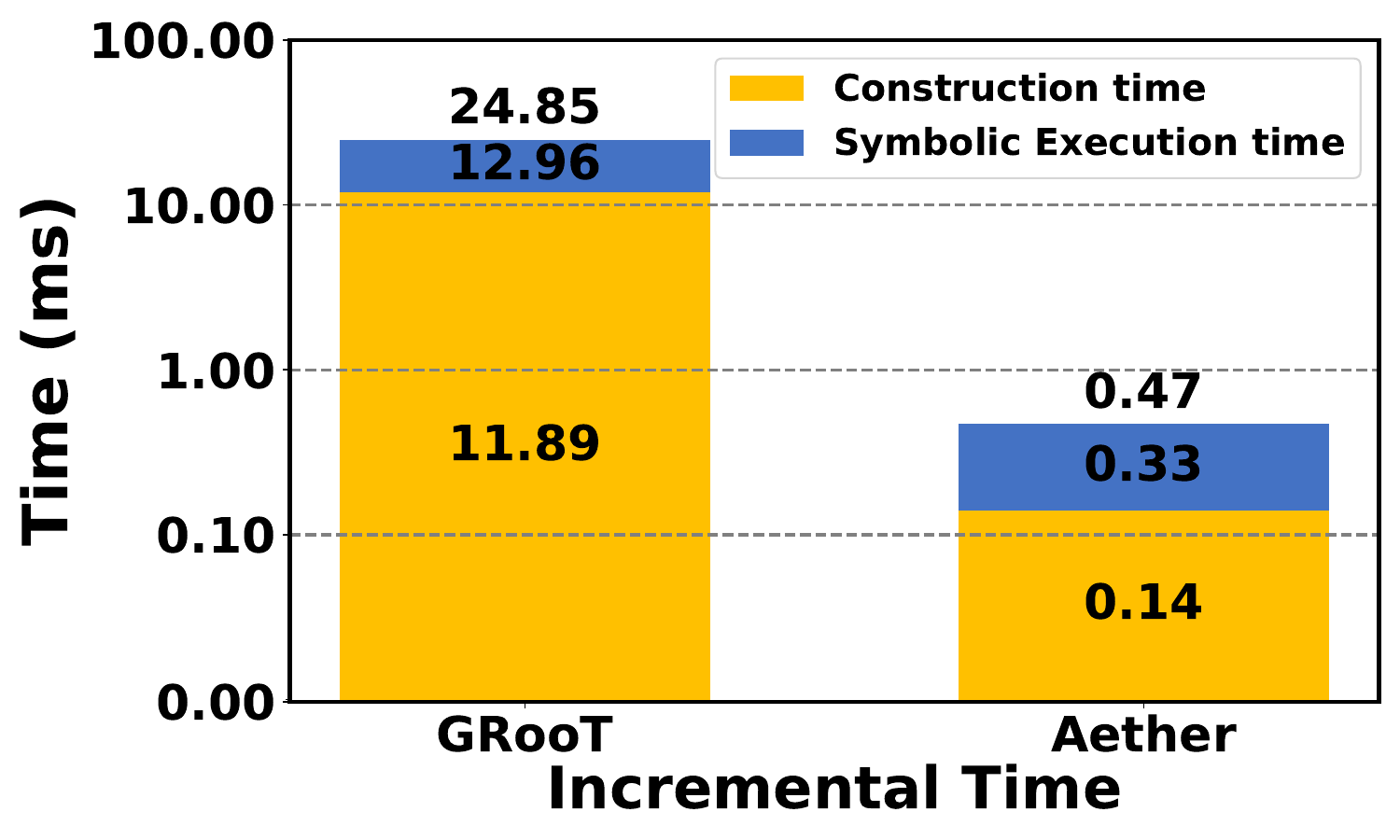}
             \caption{The mean time of end-to-end verification time.}
             \label{fig:incre_component}
        \end{subfigure}
        % \begin{subfigure}[b]{0.45\textwidth}
        %     \includegraphics[width=\textwidth]{figures/test_10_8_2.pdf}
        %     % \caption{4}
        % \end{subfigure}
    \end{minipage}\vspace{-5pt}
     \caption{Incremental update experiment result using Census dataset.}
\end{figure*}
\para{Experiment 1: Burst update.} We first evaluate \system{} in the scenario of burst update (make the complete censu dataset available to \system{} at once for full-volume evaluation). We record the number of LECs generated by \system{} on each zonefile and compare it with \groot{}. The experimental results are shown in Figure \ref{fig:burst_lecnum} indicate that for all zones, the number of ECs generated by \groot{} is greater than that produced by \system{}, with a maximum increase of 6046$\times$. For \system{}, the maximum LEC number is 96708, whereas for \groot{} is 1092168. This demonstrates that \system{} is highly effective in compressing equivalence classes. For instance, aggregating the behavior of rules containing NS records and A records means that identical resolutions for these two types of requests can be treated as a single equivalence class. In contrast, \groot{}'s compression requires enumerating these two request types separately, resulting in an exponential increase in the number of equivalence classes to compress. %\\

We also compare the construction time of LECs and ECs, and the results are shown in Figure ~\ref{fig:burst_ct}. The median construction time for \groot{} is 11.89 ms, while \system{} achieves a median of 4.65 ms. Additionally, at the 85\%, 90\%, and 95\% percentiles, \groot{} completes construction within 32.03 ms, compared to \system{}'s 11.83 ms. Furthermore, from the ratio figure the maximum speedup reaches 144.6$\times$. This is because we support parallel computation during the generation of local equivalence classes, whereas \groot{}'s architecture requires all zonfile equivalence classes to be computed sequentially.%\\
% Across 86.9\% of the zones, \system{} achieves a speedup and the maximum speedup of $144.6 \times$ in constructing LECs. Also, from the CDF of construction time (Figure~\ref{fig:burst_ct}), it is clear to see that 99.9\% quantile \system{} can finish constructing within 26.91ms, but it takes \groot{} about 203.6ms.

In symbolic execution (Figure~\ref{fig:burst_se}), the median execution time for \groot{} is 12.96 ms, while \system{} achieves a median of just 0.64 ms. At the three quantiles, the maximum time for \system{} is 1.15 ms, compared to a minimum of 22.88 ms for \groot{}. The ratio figure also indicates \system{} can achieve a maximum speedup of 2005.61$\times$. Because \system{} generates fewer equivalence classes, it reduces the need to match repetitive behaviors during symbolic execution, thereby improving the efficiency of symbolic execution.
% In symbolic execution (Figure~\ref{fig:burst_se}), the figure illustrates a significant advantage for \system{}. For 99.98\% of the data, \system{} demonstrates a shorter execution time compared to \groot{}, achieving a maximum speedup of 2005.61 times. The CDF plot shows that for the 99.9\% quantile, \system{} takes 6.32ms, while \groot{} requires 205.1ms.
Additionally, for end-to-end verification (Figure \ref{fig:burst_e2e}), \system{} achieves a median runtime of 5.29 ms, compared to 44.67 ms for \groot{}. Quantile analysis further highlights this improvement: the maximum runtime for \groot{} is 65.15 ms, whereas \system{} reduces it to just 12.54 ms. There is a maximum speedup of 255.66$\times$ of end-to-end verification, as shown in the ratio figure. And as we have mentioned befored, for 1,000,000 zones \system{} can finish verification in 140.56s that acheieves 80.87x speed up.

% Additionally, 95.4\% of the end-to-end verification cases demonstrate a speedup, with a maximum achievable speedup of 255.66$\times$. For the 99.9\% quantile, \system{} completes verification in 33.6ms, while \groot{} takes up to 50.2ms even for the 90\% quantile. This shows that \system{} has a significant effect on the overall optimization of DNS verification.
%we first evaluate \system{} in the scenario of burst update, \ie{}, all rewrite rules are rewritten to corresponding nameservers simultaneously. \system{} finishes the verification in xxxx seconds, outperforming the centralized \groot{}{} in comparison by xxx$\times$ (Figure \ref{}).

\para{Experiment 2: Incremental update.} We randomly delete and update records  in census and then evaluate the performance of \system{} and GRoot. For each zone, we randomly select a zonefile to modify according to the following rules: first, we randomly choose a record type from the rtype. We suppose the number of records in the zonefile as $num$ and calculate $Bound = \max\left(\min\left(num \times 0.01, 10\right), 1\right)$. Then, within the range of $[1, Bound]$, we generate a random number to create the corresponding number of records. Finally, we generate another random number within the range of $[1, Bound]$ to delete the specified number of records.
Throughout the running period (Figure \ref{fig:incre_combin}), \system{} consistently outperforms \groot{}, achieving a peak ratio of up to 8555.83$\times$. In symbolic execution, \system{} demonstrates speedups across all zones, with a maximum speedup of 3299.25$\times$. Similarly, in end-to-end verification, \system{} achieves speedups in all zones, reaching a maximum of 3370$\times$. Figure \ref{fig:incre_component} shows the mean of incremental end-to-end verification
time component of \system{} and \groot{}. Notably, \system{} demonstrates a speedup of 82.57$\times$, while symbolic execution achieves a speedup of 39.27$\times$. Because \system{}  only needs to update the query space affected by the modified rule, while \groot{} must re-collect the zone file, generate equivalence classes, and perform symbolic execution to support rule changes.
 \section{Conclusion}
We present \system{}, a novel DNS verification framework that achieves efficient and incremental verification through a new data structure, LEC. Experimental results demonstrate its effectiveness. In future work, we plan to investigate methods for diagnosing and repairing erroneous DNS configurations.

\emph{This work does not raise any ethical issues.}

\newpage
%\para{Acknowledgments.}
%We are extremely grateful to the anonymous SIGCOMM'25 reviewers for their feedback. We also thank Siva Kesava Reddy Kakarla, the author of {\groot}~\cite{kakarla2020groot}, for sharing the evaluation dataset. This work is supported in part by the National Key R\&D Program of China 2022YFB2901502, NSFC Award \#62172345, and NSF-Fujian-China 2022J01004.

\newpage

% \ifcase \conftype
% \bibliographystyle{ACM-Reference-Format}
% \bibliography{ref}
% \or
% \bibliographystyle{plain}
% \bibliography{ref}
% \fi

% \bibliographystyle{plainnat} % or any other style you prefer
% \bibliography{ref}
%\bibliographystyle{ACM-Reference-Format}
\bibliographystyle{plain}
\bibliography{ref}

\begin{thebibliography}{10}

\bibitem{anderson2014netkat}
Carolyn~Jane Anderson, Nate Foster, Arjun Guha, Jean-Baptiste Jeannin, Dexter Kozen, Cole Schlesinger, and David Walker.
\newblock Netkat: Semantic foundations for networks.
\newblock {\em Acm sigplan notices}, 49(1):113--126, 2014.

\bibitem{dig}
Oracle and/or~its affiliates.
\newblock dig - dns lookup utility.
\newblock \url{https://docs.oracle.com/cd/E88353_01/html/E37839/dig-1.html}, 2020.
\newblock Accessed: January 16, 2025.

\bibitem{rfc1912}
David Barr.
\newblock {Common DNS Operational and Configuration Errors}.
\newblock RFC 1912, February 1996.

\bibitem{ec1}
Conor Black and Sandra Scott-Hayward.
\newblock A survey on the verification of adversarial data planes in software-defined networks.
\newblock In {\em Proceedings of the 2021 ACM International Workshop on Software Defined Networks \& Network Function Virtualization Security}, SDN-NFV Sec'21, page 3–10, New York, NY, USA, 2021. Association for Computing Machinery.

\bibitem{BDD}
Bryant.
\newblock Graph-based algorithms for boolean function manipulation.
\newblock {\em IEEE Transactions on Computers}, C-35(8):677--691, 1986.

\bibitem{dnscensus2013}
DNS Census2013.
\newblock DNS Census2013, 2013.
\newblock Accessed: January 16, 2025.

\bibitem{Cloudfare}
Cloudfare.
\newblock 1.1.1.1 lookup failures on october 4, 2023.
\newblock \url{https://blog.cloudflare.com/1-1-1-1-lookup-failures-on-october-4th-2023}, 2023.
\newblock Accessed: January 16, 2025.

\bibitem{namedcheck}
Internet~Systems Consortium.
\newblock Linux man page: named-checkzone.
\newblock \url{https://linux.die.net/man/8/named-checkconf}, 2009.
\newblock Accessed: January 16, 2025.

\bibitem{dnsrecon}
Carlos~Perez etc.
\newblock dnsrecon.
\newblock \url{https://github.com/darkoperator/dnsrecon}, 2014.
\newblock Accessed: January 16, 2025.

\bibitem{miscase1}
Incident~Report for npm.
\newblock Dns misconfiguration cached in isp dns caches.
\newblock \url{https://status.npmjs.org/incidents/v22ffls5cd6h}, 2018.
\newblock Accessed: January 16, 2025.

\bibitem{miscase2}
James Fryman.
\newblock Dns outage post mortem.
\newblock \url{https://github.blog/2014-01-18-dns-outage-post-mortem}, 2014.
\newblock Accessed: January 16, 2025.

\bibitem{flash}
Dong Guo, Shenshen Chen, Kai Gao, Qiao Xiang, Ying Zhang, and Y.~Richard Yang.
\newblock Flash: fast, consistent data plane verification for large-scale network settings.
\newblock In {\em Proceedings of the ACM SIGCOMM 2022 Conference}, SIGCOMM '22, page 314–335, New York, NY, USA, 2022. Association for Computing Machinery.

\bibitem{ec3}
Arpit Gupta, Laurent Vanbever, Muhammad Shahbaz, Sean~P. Donovan, Brandon Schlinker, Nick Feamster, Jennifer Rexford, Scott Shenker, Russ Clark, and Ethan Katz-Bassett.
\newblock Sdx: a software defined internet exchange.
\newblock In {\em Proceedings of the 2014 ACM Conference on SIGCOMM}, SIGCOMM '14, page 551–562, New York, NY, USA, 2014. Association for Computing Machinery.

\bibitem{ec2}
Sylvain Hall\'{e}.
\newblock Test suite generation for boolean conditions with equivalence class partitioning.
\newblock In {\em Proceedings of the IEEE/ACM 10th International Conference on Formal Methods in Software Engineering}, FormaliSE '22, page 23–33, New York, NY, USA, 2022. Association for Computing Machinery.

\bibitem{rfc8484}
Paul~E. Hoffman and Patrick McManus.
\newblock {DNS Queries over HTTPS (DoH)}.
\newblock RFC 8484, 2018.

\bibitem{checkhost2020}
Check Host.
\newblock Check host.
\newblock \url{http://check-host.net/check-dns}, 2020.
\newblock Accessed: January 16, 2025.

\bibitem{oxidd24}
Nils Husung, Clemens Dubslaff, Holger Hermanns, and Maximilian~A. K{\"o}hl.
\newblock {OxiDD}: A safe, concurrent, modular, and performant decision diagram framework in {Rust}.
\newblock In {\em Proceedings of the 30th International Conference on Tools and Algorithms for the Construction and Analysis of Systems (TACAS'24)}, 2024.

\bibitem{miscase3}
InfinityFree.
\newblock Dns outage at ifastnet: Softaculous down.
\newblock \url{https://forum.infinityfree.com/t/dns-outage-at-ifastnet-softaculous-down/19374}, 2019.
\newblock Accessed: January 16, 2025.

\bibitem{bind9}
{Internet Systems Consortium, Inc.}
\newblock {BIND - DNS Software}.
\newblock \url{https://www.isc.org/bind/}, 2022.
\newblock Accessed: January 16, 2025.

\bibitem{kakarla2020groot}
Siva Kesava~Reddy Kakarla, Ryan Beckett, Behnaz Arzani, Todd Millstein, and George Varghese.
\newblock Groot: Proactive verification of dns configurations.
\newblock In {\em SIGCOMM'20}, pages ACM, 310--328, 2020.

\bibitem{veriflow}
Ahmed Khurshid, Wenxuan Zhou, Matthew Caesar, and P.~Brighten Godfrey.
\newblock Veriflow: verifying network-wide invariants in real time.
\newblock 42(4):467–472, September 2012.

\bibitem{rfc3467}
J.~Klensin.
\newblock Rfc3467: Role of the domain name system (dns), 2003.

\bibitem{liu2023formal}
Si~Liu, Huayi Duan, Lukas Heimes, Marco Bearzi, Jodok Vieli, David Basin, and Adrian Perrig.
\newblock A formal framework for end-to-end dns resolution.
\newblock In {\em SIGCOMM'23}, pages ACM, 932--949, 2023.

\bibitem{microsoft2020dnslint}
Microsoft.
\newblock Description of the dnslint utility, 2020.
\newblock Accessed: January 16, 2025.

\bibitem{rfc1034}
P.~Mockapetris.
\newblock {Domain names - concepts and facilities}.
\newblock RFC 1034, 1987.

\bibitem{rfc1035}
P.~Mockapetris.
\newblock {Domain names - implementation and specification}.
\newblock RFC 1035, 1987.

\bibitem{dnsdev}
Paul Mockapetris and Kevin~J Dunlap.
\newblock Development of the domain name system.
\newblock In {\em Symposium proceedings on Communications architectures and protocols}, pages 123--133, 1988.

\bibitem{Roll}
Moritz M\"{u}ller, Matthew Thomas, Duane Wessels, Wes Hardaker, Taejoong Chung, Willem Toorop, and Roland~van Rijswijk-Deij.
\newblock Roll, roll, roll your root: A comprehensive analysis of the first ever dnssec root ksk rollover.
\newblock In {\em Proceedings of the Internet Measurement Conference}, IMC '19, page 1–14, New York, NY, USA, 2019. Association for Computing Machinery.

\bibitem{MXToolbox}
Inc MXToolBox.
\newblock Mxtoolbox.
\newblock \url{https://mxtoolbox.com}, 2004.
\newblock Accessed: January 16, 2025.

\bibitem{pang2004}
Jeffrey Pang, Aditya Akella, Anees Shaikh, Balachander Krishnamurthy, and Srinivasan Seshan.
\newblock On the responsiveness of dns-based network control.
\newblock In {\em Proceedings of the 4th ACM SIGCOMM Conference on Internet Measurement}, IMC '04. Association for Computing Machinery, 2004.

\bibitem{ec6}
Federico Parola, Roberto Procopio, Roberto Querio, and Fulvio Risso.
\newblock Comparing user space and in-kernel packet processing for edge data centers.
\newblock {\em SIGCOMM Comput. Commun. Rev.}, 53(1):14–29, April 2023.

\bibitem{bobcares2020}
Keerthi PS.
\newblock Bind9 check zone file - how we do it, Mar 2020.
\newblock Accessed: January 16, 2025.

\bibitem{schomp2013}
Kyle Schomp, Tom Callahan, Michael Rabinovich, and Mark Allman.
\newblock On measuring the client-side dns infrastructure.
\newblock In {\em Proceedings of the 2013 Conference on Internet Measurement Conference}, IMC '13. Association for Computing Machinery, 2013.

\bibitem{ec5}
Colin Scott, Andreas Wundsam, Barath Raghavan, Aurojit Panda, Andrew Or, Jefferson Lai, Eugene Huang, Zhi Liu, Ahmed El-Hassany, Sam Whitlock, H.B. Acharya, Kyriakos Zarifis, and Scott Shenker.
\newblock Troubleshooting blackbox sdn control software with minimal causal sequences.
\newblock In {\em Proceedings of the 2014 ACM Conference on SIGCOMM}, SIGCOMM '14, page 395–406, New York, NY, USA, 2014. Association for Computing Machinery.

\bibitem{shaikh2001}
A.~Shaikh, R.~Tewari, and M.~Agrawal.
\newblock On the effectiveness of dns-based server selection.
\newblock In {\em Proceedings IEEE INFOCOM 2001. Conference on Computer Communications. Twentieth Annual Joint Conference of the IEEE Computer and Communications Society (Cat. No.01CH37213)}, 2001.

\bibitem{miscase5}
Liam Tung.
\newblock Azure global outage: Our dns update mangled domain records, says microsoft.
\newblock \url{https://www.zdnet.com/article/azure-global-outage-our-dns-update-mangled-domain-records-says-microsoft/}, 2019.
\newblock Accessed: January 16, 2025.

\bibitem{ec4}
Hao Wang, Haiyong Xie, Lili Qiu, Yang~Richard Yang, Yin Zhang, and Albert Greenberg.
\newblock Cope: traffic engineering in dynamic networks.
\newblock {\em SIGCOMM Comput. Commun. Rev.}, 36(4):99–110, August 2006.

\bibitem{Octopus}
Yao Wang, Kexin Yu, Ziyi Wang, Kaiqiang Hu, Haizhou Du, Qiao Xiang, Xing Fang, Geng Li, Ruiting Zhou, Linghe Kong, and Jiwu Shu.
\newblock Rethinking dns configuration verification with a distributed architecture.
\newblock In {\em Proceedings of the 8th Asia-Pacific Workshop on Networking}, APNet '24, page 23–30, New York, NY, USA, 2024. Association for Computing Machinery.

\bibitem{miscase4}
Zack Whittaker.
\newblock A dns outage just took down a large chunk of the internet.
\newblock \url{https://techcrunch.com/2021/07/22/a-dns-outage-just-took-down-a-good-chunk-of-the-internet/}, 2021.
\newblock Accessed: January 16, 2025.

\bibitem{xu_2015_systems}
Tianyin Xu and Yuanyuan Zhou.
\newblock Systems approaches to tackling configuration errors: A survey.
\newblock {\em ACM Computing Surveys}, 47(4):1--41, 2015.

\bibitem{monitoring_zdrnja}
Bojan Zdrnja, Nevil Brownlee, and Duane Wessels.
\newblock Passive monitoring of dns anomalies.
\newblock In {\em International Conference on Detection of Intrusions and Malware, and Vulnerability Assessment}, pages 129--139. Springer, 2007.

\end{thebibliography}

\appendix
\newpage

\newpage
\textbf{Appendices are supporting material that has not been peer-reviewed.} 
\section{Pseudocode for constructing LEC for a zonefile. }
Here is the pseudocode th demonstrates how \system{} constructs the LEC for a single zonefile in detail. 
\begin{algorithm}
    \footnotesize
    \caption{Construct the LECs for a zonefile}\label{algleczone}

    \SetKwFunction{GetSpace}{GetSpace}

    \SetKwFunction{Rank}{Rank}
    \SetKwProg{Fn}{Function}{:}{}
    \Fn{\Rank{r, origin}}{
        \If{$r.rtype = \bf{NS}$ \textbf{and} $r.rname \neq origin$}{
            \KwRet{$512 - r.rname.num\_labels * 2$}\;
        }
        \ElseIf{$r.rtype = \bf{DNAME}$}{
            \KwRet{$512 - r.rname.num\_labels * 2 - 1$}\;
        }
        \Else{
            $is\_cname \leftarrow r.rtype = \bf{CNAME}$\;
            $is\_not\_wildcard \leftarrow * \notin r.rname$\;
            \KwRet{$is\_not\_wildcard * 2 + is\_cname$}\;
        }
    }
    
    \SetKwFunction{ZoneLECs}{ZoneLECs}
    \SetKwProg{Fn}{Function}{:}{}
    \Fn{\ZoneLECs{zonefile, $remain\_bdd$}}{
        Aggregate $records$ with the same rname and rtype\;
        $records \leftarrow empty\ list$\;
        \ForEach{record r $\langle rname,rtype,rdata\rangle$ in zonefile}{
            $rank \leftarrow $ \Rank{r, zonefile.origin}\;
            $records.push(\langle rname,rtype,rdata,rank\rangle)$\;
            \If{$r.rtype = \bf{DNAME}$}{
                $records.push(\langle rname,rtype,rdata,2\rangle)$\;
            }
            \ElseIf{$r.rtype = \bf{CNAME}$}{
                $records.push(\langle rname,rtype,rdata,rank-1\rangle)$\;
            }
        }
        Sort $records$ by decreasing rank\;
        $rules \leftarrow empty\ list$\;
        \ForEach{$\langle rname,rtype,rdata,rank\rangle$ in records}{
            \If{$rank > 3$ \textbf{and} $rank\ \%\ 2 = 0$}{
                $rule\_hit \leftarrow $ \GetSpace{rname, $\bf{all\_types}$, 1}\;
                $action \leftarrow \langle Delegate,adata\rangle$\;
            }
            \ElseIf{$rank > 3$ \textbf{and} $rank\ \%\ 2 = 1$}{
                $rule\_hit \leftarrow $ \GetSpace{rname, $\bf{all\_types}$, 2}\;
                $action \leftarrow \langle RewriteD,adata\rangle$\;
            }
            \ElseIf{$rank = 1$ \textbf{or} $rank = 3$}{
                $rule\_hit \leftarrow $ \GetSpace{rname, $\bf{all\_types}$ - $\bf{CNAME}$, 0}\;
                $action \leftarrow \langle RewriteC,adata\rangle$\;
            }
            \Else{
                $rule\_hit \leftarrow $ \GetSpace{rname, rtype, 0}\;
                $action \leftarrow \langle Answer,adata\rangle$\;
            }
            $rule\_bdd \leftarrow rule\_hit\cap remain\_bdd$\;
            $remain\_bdd \leftarrow remain\_bdd - rule\_bdd$\;
            $rules.push(\langle rule\_hit,rule\_bdd, action \rangle)$\;
        }
        
        \KwRet{(rules, $\langle remain\_bdd,remain\_bdd,\langle NonExist,()\rangle\rangle$)}\;
    }

\end{algorithm}
\\
\\

%\printbibliography

\end{document}